\documentstyle[12pt,epsf]{article}
\evensidemargin\oddsidemargin
 1
 1
 1

\newcommand{\alt}{\mathrel{\raisebox{-.6ex}{$\stackrel{\textstyle<}{\sim}$}}}
\newcommand{\agt}{\mathrel{\raisebox{-.6ex}{$\stackrel{\textstyle>}{\sim}$}}}

\def\overlay#1#2{\ifmmode \setbox 0=\hbox {$#1$}\setbox 1=\hbox to\wd 0{\hss
$#2$\hss }\else \setbox 0=\hbox {#1}\setbox 1=\hbox to\wd 0{\hss #2\hss }\fi
#1\hskip -\wd 0\box 1}

\catcode`@=11
\newtoks\@stequation

\def\mathletters{\refstepcounter{equation}%
  \edef\@savedequation{\the\c@equation}%
  \@stequation=\expandafter{\theequation}
  \edef\@savedtheequation{\the\@stequation}
  \edef\oldtheequation{\theequation}%
  \setcounter{equation}{0}%
  \def\theequation{\oldtheequation\alph{equation}}}

\def\endmathletters{%
  \setcounter{equation}{\@savedequation}%
  \@stequation=\expandafter{\@savedtheequation}%
  \edef\theequation{\the\@stequation}%
  \global\@ignoretrue}
\catcode`=12

\begin{document}

\font\fortssbx=cmssbx10 scaled \magstep2
\hbox to \hsize{
\hskip.5in \raise.1in\hbox{\fortssbx University of Wisconsin - Madison}
\hfill\vbox{\hbox{\bf MAD/PH/801}
            \hbox{November 1993}} }

\baselineskip28pt

\begin{center}
{\Large\bf   THE SUPERSYMMETRIC PARTICLE SPECTRUM}\\[.2in]
{\large V.~Barger,  M.S.~Berger, and P.~Ohmann}\\[.1in]
\it
Physics Department, University of Wisconsin, Madison, WI 53706, USA
\end{center}

\renewcommand{\LARGE}{\Large}
\renewcommand{\Huge}{\Large}

\begin{abstract}
We examine the spectrum of supersymmetric particles predicted by
grand unified theoretical (GUT) models where
the electroweak symmetry breaking is accomplished radiatively. We evolve the
soft supersymmetry breaking parameters according to the renormalization group
equations (RGE).
The minimization of the Higgs potential is conveniently described
by means of tadpole diagrams.
We present complete one-loop expressions
for these minimization conditions,
including contributions
from the matter and the gauge sectors.
We concentrate on the low $\tan \beta$ fixed point region
(that provides a natural explanation of a large top quark mass)
for which we find solutions to the RGE satisfying both experimental bounds
and fine-tuning criteria.
We also find that the constraint from the consideration of the lightest
supersymmetric particle as the dark matter of the universe is
accommodated in much of parameter space
where the lightest neutralino is predominantly
gaugino. The supersymmetric mass spectrum displays correlations that are
model-independent over much of the GUT parameter space.
\end{abstract}

\section{Introduction}
Why should one be interested in supersymmetry?
Until recently, the reasons have been principally theoretical.
Supersymmetry (SUSY) is a beautiful extension of the Poincar\'{e} symmetry
with new dimensions of space and time that explain the existence
of fermions\cite{susy}. It solves the hierarchy problem of widely separated
electroweak and grand unified scales through cancellations among
diagrams that give quadratically divergent Higgs boson mass
corrections. Moreover supersymmetry may be a necessary consequence
of string theory.

The recent upswing in interest in supersymmetry derives from high
precision measurements of Standard Model (SM) parameters at LEP.
Renormalization group evolution with minimal SM particle
content of the SU(3), SU(2), and U(1)
couplings from $Q^2 = {M_Z}^2$ do not converge at a single high
scale, in contradiction with the prediction of the SU(5) grand
unified theory (GUT). However, with the minimal particle content
of supersymmetry included, the evolution is in excellent agreement with
LEP data and suggests a grand unified scale at $M_G \simeq 2\times 10^{16}$ GeV
and effective SUSY mass scale within the range
$M_Z^{} <  M_{SUSY}^{}< 1$
TeV\cite{susygut}. Encouraged by this success, the evolution of Yukawa
couplings is also being vigorously pursued, with Yukawa unification
constraints such as $\lambda _b=\lambda _{\tau}$ at the GUT scale\cite{ceg}.
While the unification of gauge and Yukawa couplings is an extremely
attractive feature, the existence of supersymmetry will only be confirmed
when new particle states are seen directly and the associated
R-parity conservation or violation is tested in the production and decays of
these supersymmetric particles.

The idea of a radiative breaking of the electroweak symmetry is an old but
still popular one\cite{nac}--\cite{copw}. It is very attractive to explain the
breaking of the electroweak symmetry through large logarithms between the
Planck scale and the weak scale\cite{radiative}.
For the radiative corrections to be
strong enough to drive a Higgs boson mass-squared parameter negative
(thus breaking the electroweak symmetry), a Yukawa coupling of that
Higgs boson must be large at the GUT scale. With
the top quark mass large ($m_t > 100$ GeV), the SUSY GUT
unification can naturally explain the origin of the electroweak scale.
A heavy top is required to drive one of
the soft-supersymmetry breaking parameters (a Higgs doublet mass) negative.
Today we know the top quark mass is
large and that the top has a large Yukawa coupling.
There is a relationship between the electroweak scale and the
top quark Yukawa coupling through the RGEs; consequently
the radiative symmetry breaking
mechanism has important consequences for the
supersymmetric particle spectrum. Indeed a large top Yukawa coupling is the
motivation for the fixed point solutions\cite{pendleton} advocated
recently in the context of GUT theories\cite{bbo}--\cite{BCPW}.
These solutions predict a linear relationship
between $m_t$ and $\sin \beta $, given further constraints on  the SUSY
particle spectrum.

There are at least two other motivations for supersymmetry.
In the context of SUSY GUTs, the grand unification scale is raised
sufficiently high to suppress
proton decay to experimentally acceptable levels, when
an additional R-parity symmetry is invoked.
R-parity symmetry
has an important consequence, providing the second
additional motivation for supersymmetry -- it implies that the lightest
supersymmetric particle (LSP) is stable.
It is now generally believed that baryonic matter
is insufficient to make up the total gravitationally interacting
matter of the universe.
The LSP
provides a natural candidate for the (cold) dark matter of the
universe, since the LSP is forbidden to decay into baryons by R-parity
conservation.

The greatest dilemma for supersymmetry is the unknown mechanism for
supersymmetry breaking. Since no supersymmetric partners of the known particles
have been observed, one must not only explain why supersymmetry is
broken, but also why it is broken in such a way that (almost) all
of the supersymmetric partners are heavier than the known particles.
Typically one skirts the issue of the exact mechanism of supersymmetry
breaking, and parametrizes one's ignorance by introducing soft-supersymmetry
breaking parameters which characterize the scale of the masses of the
supersymmetric particles. In the context of certain models (minimal
supergravity, no-scale supergravity, superstring-inspired, etc.) there are
relations between these parameters at the GUT scale
that appear as universal boundary
conditions. The values of the parameters would be determined in some still
unknown fundamental theory, but in practice the models are described as
depending on a few undetermined soft-supersymmetry breaking masses which,
when evolved in the well-known effective theory below the unification scale,
generate consequences for the masses and couplings of the SUSY particles
that can be compared to experiment.

The supersymmetry breaking in such models occurs through a
hidden sector at some
scale intermediate between the electroweak and Planck scales.
For example, there may exist
an asymptotically free gauge theory in the hidden sector
in which gaugino condensation or some other nonperturbative mechanism
occurs at a high scale $H_{SUSY}^{}$ related
by a geometric hierarchy to the Planck and electroweak scales
($M_{SUSY}^{}\sim H_{SUSY}^n/M_P^{n-1}$), thus breaking supersymmetry
in the observable sector near the electroweak scale $M_{SUSY}^{}\sim M_W^{}$.
Since this breaking of supersymmetry occurs in the hidden sector which is
coupled only gravitationally to the observable sector, the supersymmetry
breaking is suppressed relative to the Planck scale by factors of
$H_{SUSY}^{}/M_P^{}$. This mechanism gives rise to an effective
theory in the observable sector with softly broken global
supersymmetry, giving the
scale of the soft-supersymmetry breaking parameters near the electroweak scale.
How the effective theory parameters are determined and related to each other
depends on the details of the supersymmetry breaking scenario.

\section{Soft-supersymmetry breaking parameters}

Retaining only the dominant Yukawa couplings $\lambda _t$, $\lambda _b$ and
$\lambda _\tau$, the superpotential\footnote{Caution: Our convention
for the sign of $\mu $
(and also the sign of the trilinear couplings $A$ introduced below)
differs from some authors.}
is given in terms of the superfields by
\begin{eqnarray}
W&=&\epsilon _{ij}\left (\lambda _tQ^iH_2^jt^c+\lambda _bQ^iH_1^jb^c
+\lambda _\tau L^iH_1^j\tau ^c+\mu H_1^iH_2^j\right )\;,
\end{eqnarray}
where $Q=(t,b)$, $L=(\tau , \nu _\tau)$ and $H_1=(H_1^0,H_1^-)$ and
$H_1=(H_2^+,H_2^0)$ and $\epsilon _{ij}$ with $i$, $j=1$, $2$
is the antisymmetric tensor in two dimensions.
The Yukawa couplings are defined by
\begin{eqnarray}
\lambda _t={{\sqrt{2}m_t}\over {v\sin \beta}}\;, \qquad
\lambda _b={{\sqrt{2}m_b}\over {v\cos \beta}}\;, \qquad
\lambda _{\tau }={{\sqrt{2}m_{\tau}}\over {v\cos \beta}}\;,
\end{eqnarray}
where $\tan \beta =v_2/v_1$ is the ratio of the vacuum expectation values
of $H_2^0$ and $H_1^0$.
The $\mu $ term in the superpotential contributes to the Higgs potential which
at tree level is
\begin{eqnarray}
V_0&=&(m_{H_1}^2+\mu ^2)|H_1|^2+(m_{H_2}^2+\mu
^2)|H_2|^2+m_3^2(\epsilon_{ij}{H_1}^i{H_2}^j+{\rm h.c.})
\nonumber \\
&&+{1\over 8}(g^2+g^{\prime 2})\left [|H_1|^2-|H_2|^2\right ]^2
+{1\over 2}g^2|H_1^{i*}H_2^i|^2\;, \label{tree}
\end{eqnarray}
where $m_{H_1}^{}$, $m_{H_2}^{}$, and
$m_3$ are soft-supersymmetry breaking parameters. We shall
define as usual the soft Higgs mass parameters
\begin{mathletters}
\begin{eqnarray}
m_1^2&=&m_{H_1}^2+\mu ^2\;, \\
m_2^2&=&m_{H_2}^2+\mu ^2\;.
\end{eqnarray}
\end{mathletters}
Of the eight degrees of
freedom in the two Higgs doublets, three ($G^{\pm }$, $G^0$) are absorbed to
give the $W^{\pm }$ and $Z$ masses, leaving five physical Higgs bosons:
the charged Higgs bosons $H^{\pm }$, the CP-even Higgs bosons $h$ and $H$, and
the CP-odd Higgs boson $A$.

There are soft-supersymmetry breaking gaugino mass terms
\begin{eqnarray}
{1\over 2}M_1\overline{B}B+{1\over 2}M_2\overline{W}^aW^a
+{1\over 2}M_3\overline{\tilde{g}^b}\tilde{g}^b\;,
\end{eqnarray}
for the bino $B$, the winos $W^a$ ($a=1$, $2$, $3$), and the gluinos
$\tilde{g}^b$
($b=1,\dots ,8$). Corresponding to each superpotential coupling
there is a soft-supersymmetry breaking trilinear coupling
\begin{eqnarray}
\epsilon _{ij}\left (\lambda _tA_t\tilde {Q}^iH_2^j\tilde{t}^c
+\lambda _bA_b\tilde{Q}^iH_1^j\tilde{b}^c
+\lambda _{\tau }A_{\tau}\tilde{L}^iH_1^j\tilde{\tau }^c
+\mu BH_1^iH_2^j\right )
\end{eqnarray}
and soft squark and slepton mass terms
\begin{eqnarray}
&&M_Q^2[\tilde{t}_L^*\tilde{t}_L+\tilde{b}_L^*\tilde{b}_L]
+M_U^2\tilde{t}_R^*\tilde{t}_R+M_D^2\tilde{b}_R^*\tilde{b}_R\nonumber \\
&+&M_L^2[\tilde{\tau }_L^*\tilde{\tau }_L+\tilde{\nu}_L^*\tilde{\nu}_L]
+M_E^2\tilde{\tau }_R^*\tilde{\tau }_R\;.
\end{eqnarray}
The RGE for the soft-supersymmetry breaking parameters are given in the
appendix, and the RGE for the gauge and Yukawa couplings are summarized
in Ref.~\cite{bbo}.

An interesting aspect of the supergravity breaking mechanism is the origin of
the $3-2-1$ supersymmetry at low scales. Why is it the electroweak gauge
group the one that is broken,
and not QCD? Consider the renormalization group equations from the
appendix for the scalar states $H_2$, $\tilde{t}_R$, and $\tilde{Q}_L$
retaining only the QCD gauge coupling $g_3$ and the
top Yukawa coupling $\lambda _t$ terms\cite{hall},
\begin{eqnarray}
8\pi ^2{d\over {dt}}\left \{\begin{array}{c}
M_{H_2}^2 \\
M_{t_R}^2 \\
M_{Q_L}^2
\end{array}\right \}&=&
-{{16}\over 3}g_3^2M_3^2\left \{\begin{array}{c}
0 \\
1 \\
1
\end{array}\right \}+\lambda _t^2X_t\left \{\begin{array}{c}
3 \\
2 \\
1
\end{array}\right \}\;,
\end{eqnarray}
where $X_t=M_{Q_L}^2+M_{t _R}^2+M_{H_2}^2+A_t^2$ and $t=\ln Q/M_G$.
The $\lambda _t^2$ term is the means by which the mass-squares are
driven to lower values as the scale decreases. Because
the Higgs field is uncolored, the group theory factors
allow $M_{H_2}^2$ to be driven negative with $M_{t_R}^2$ and
$M_{Q_L}^2$ remaining positive, thus breaking only the electroweak gauge group.

According to
conventional wisdom the squarks and sleptons have a universal
soft-supersymmetry breaking mass $m_0^{}$ at the unification scale. Then
any deviations from degeneracy at the SUSY scale
are suppressed by the associated quark or
lepton mass, which is small except for the top squarks. The
flavor changing neutral currents (FCNCs) are thereby
suppressed to an acceptable level.
The universal boundary condition applies in minimal supergravity models
with the canonical kinetic energy. Recently there has been some interest
in relaxing this condition\cite{dilaton}--\cite{gr}.

Analytical expressions can be obtained for the squark and slepton
mass parameters when
the corresponding Yukawa couplings are negligible (i.e. for the first two
generations). For a universal scalar mass
$m_0^{}$ and gaugino mass $m_{1 \over 2}^{}$
at the GUT scale (this condition
need not apply in general in string theories), one has the relation
\begin{eqnarray}
m_{\tilde {f}}^2&=&m_0^2+\sum _{i=1}^3f_im_{1\over 2}^2+(T_{3,\tilde{f}}
-e_{\tilde{f}}\sin ^2\theta _w)M_Z^2\cos
2\beta \;, \label{fi}
\end{eqnarray}
for the squark and slepton masses
where the $f_i$ are (positive) constants that depend on the evolution of the
gauge couplings
\begin{eqnarray}
f_i&=&{{c_i({f})}\over {b_i}}\left [1-{1\over
{\left ({1-{{\alpha _G^{}}\over {2\pi }}b_it}\right )^2}}\right ]\;.
\end{eqnarray}
Here $T_{3,\tilde{f}}$ is the SU(2) quantum number and
$e_{\tilde{f}}$ is the electromagnetic charge of the sfermion.
The $b_i$ are given in the appendix and $c_i({f})$ is
${{N^2-1}\over {N}}$ $(0)$ for fundamental (singlet) representations
of $SU(N)$ and ${3\over {10}}Y^2$ for $U(1)_Y^{}$.
The squark mass spectrum of the
third generation is more complicated for two reasons: (1) the effects of the
third generation Yukawa couplings need not be negligible, and (2) there can
be substantial mixing between the left and right top squark fields
(and left and right bottom squark fields for large $\tan\beta$) so that
they are not the mass eigenstates.

The gaugino evolution is particularly simple by virtue of their simple
renormalization group equations; at one-loop order
the gaugino masses parameters $M_1$, $M_2$, and
$M_3$ scale in exactly
the same proportions as do the gauge couplings so that
\begin{eqnarray}
m_{\tilde{g}}&=&M_3(m_t)={\alpha_3(m_t) \over \alpha_2(m_t)}M_2(m_t)
={\alpha_3(m_t) \over \alpha_1(m_t)}M_1(m_t)\;.
\end{eqnarray}

Figure 1 shows a typical evolution of the soft-supersymmetry
breaking parameters. The characteristic
behavior exhibited by the mass parameters are typical
of renormalization group equation evolution. The colored particles
are generally driven heavier at low $Q$
by the large strong gauge coupling. The
Higgs mass parameter $m_2^2$ is usually driven negative
(at least for $\tan\beta$ not too small), giving the electroweak
symmetry breaking. Assumed universal boundary conditions at the GUT scale
yields correlations between the masses in the supersymmetric spectrum.

\begin{center}
\epsfxsize=6.25in
\hspace*{0in}
\epsffile{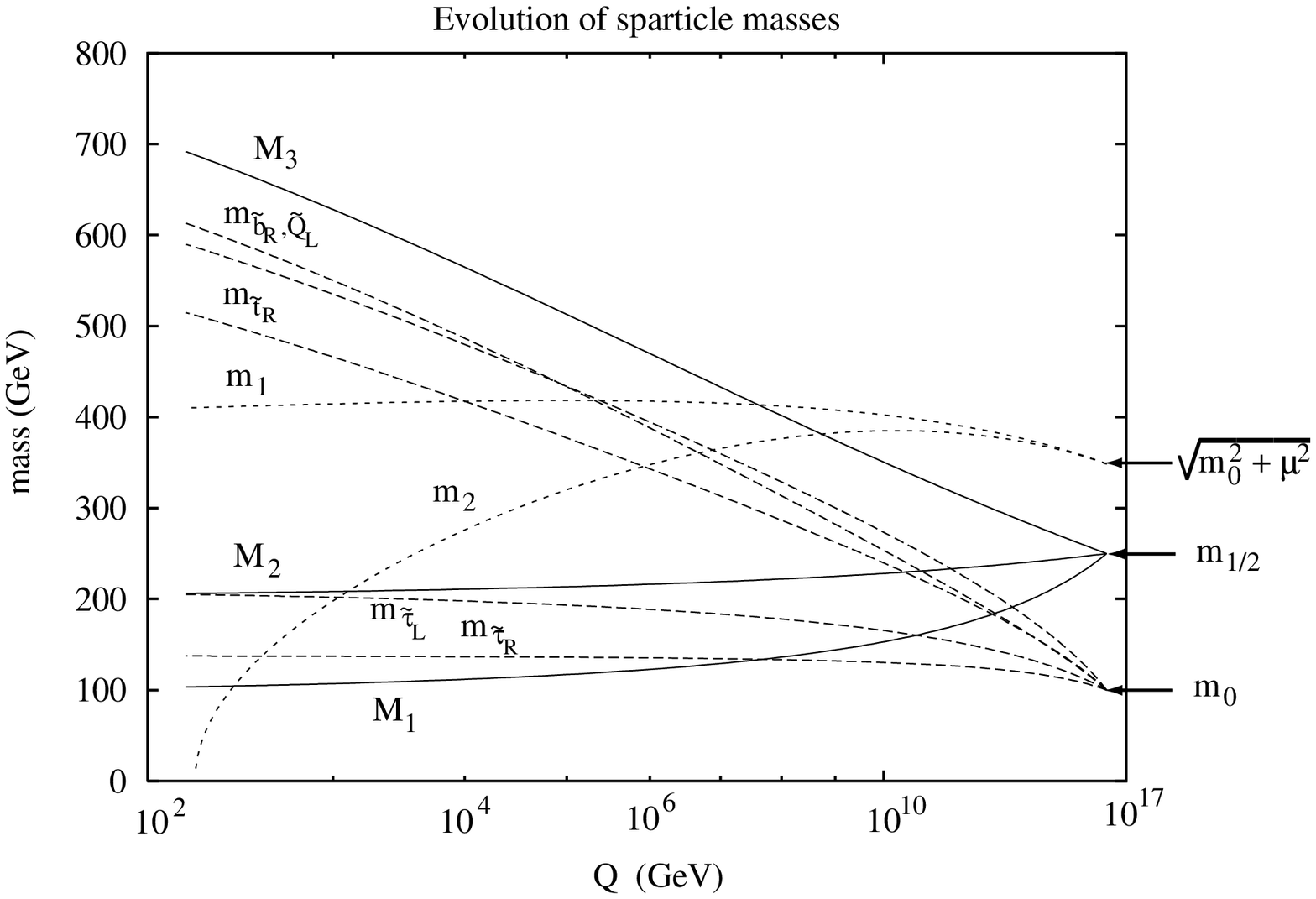}

\parbox{5.5in}{\small Fig.~1. An example of the running of the
soft-supersymmetry breaking parameters for
$\alpha_s(M_Z) = 0.120$,
$m_t(m_t) = 150$ GeV, $\tan \beta = 10$, $m_{1 \over 2} = 250$ GeV, $m_0 = 100$
GeV,
and $A^G = 0$, where the superscript $G$ denotes the GUT scale.\\}
\end{center}

Fixed-point solutions to the RGE predict that
the scale of the top-quark mass is naturally large in SUSY-GUT models but
depends on $\tan\beta$. The prediction is that\cite{bbo}
\begin{eqnarray}
m_t^{\rm pole}&=&(200\;{\rm GeV})\sin \beta\;. \label{fixedpt}
\end{eqnarray}
Note that the propagator-pole mass $m_t^{\rm pole }$ is related to this
running mass $m_t(m_t)$ by\cite{Tarrach}
\begin{equation}
m_t^{\rm pole} = m_t(m_t)\left[1+{4\over3\pi}\alpha_3(m_t)
\right]\,.
\label{mtpole}
\end{equation}

\section{One-Loop Contributions: Tadpole Method}

Although the tree level Higgs potential is not reliable for the
purpose of analyzing
radiative breaking of the electroweak symmetry\cite{grz},
it provides a convenient starting point for our discussion.
Recall the tree level potential Eq.~(\ref{tree}).
In supergravity models $m_3^2$ is related to $B$ and $\mu $ by
\begin{eqnarray}
m_3^2&=&B\mu \;.
\end{eqnarray}
When the neutral components of the Higgs doublets receive vacuum expectation
values $v_1$ and $v_2$, the potential develops tadpoles. Inserting\cite{dh}
\begin{mathletters}
\begin{equation}
H_1=\left( \begin{array}{c}
{1\over {\sqrt{2}}}(\psi_1+v_1+i\phi_1) \\
H_1^-
\end{array} \right) \label{doublet1}
\end{equation}
\begin{equation}
H_2=\left( \begin{array}{c}
H_2^+ \\
{1\over {\sqrt{2}}}(\psi_2+v_2+i\phi_2)
\end{array} \right)  \label{doublet2}
\end{equation}
\end{mathletters}
into Eq.~(\ref{tree}) one can identify
\begin{eqnarray}
V_{\rm tadpole}&=&t_1\psi _1+t_2\psi _2\;,
\end{eqnarray}
where $t_1$ and $t_2$ are (tree-level) tadpoles:
\begin{mathletters}
\begin{eqnarray}
t_1&=&(m_{H_1}^2+\mu ^2)v_1+B\mu v_2
+{1\over 8}(g^2+g^{\prime 2})v_1(v_1^2-v_2^2)
\;, \label{tadpole1} \\
t_2&=&(m_{H_2}^2+\mu ^2)v_2+B\mu v_1
-{1\over 8}(g^2+g^{\prime 2})v_2(v_1^2-v_2^2)
\;. \label{tadpole2}
\end{eqnarray}
\end{mathletters}
The minimum of the Higgs potential is determined by setting the first
derivatives of the fields to zero,
\begin{eqnarray}
{{\partial V_0}\over {\partial \psi_i }}=
{{\partial V_{tadpole}}\over {\partial \psi_i }}
=0\;. \label{Vmintree}
\end{eqnarray}
Therefore the tadpoles $t_1$ and $t_2$ must vanish at the minimum.
With our normalization of $\psi_1$ and $\psi_2$
(i.e. including the factor of ${1\over {\sqrt{2}}}$ in
Eq.~(\ref{doublet1}) and Eq.~(\ref{doublet2})), the $W$ and $Z$ masses are
\begin{mathletters}
\begin{eqnarray}
M_W^2&=&{1\over 4}g^2(v_1^2+v_2^2)
\;, \label{mwtree} \\
M_Z^2&=&{1\over 4}(g^2+g^{\prime 2})(v_1^2+v_2^2)
\;, \label{mztree}
\end{eqnarray}
\end{mathletters}
which implies $v_1^2+v_2^2=v^2=(246\rm{ GeV})^2$.
A particularly useful form of the minimization conditions is obtained
by forming the linear combinations $T_1$ and $T_2$ of the tadpoles given by
\begin{equation}
\left( \begin{array}{c}
T_1 \\ T_2
\end{array} \right)= \left( \begin{array}{c@{\quad}c}
\cos \beta  & -\sin \beta  \\
\sin \beta  & \cos \beta
\end{array} \right)\left( \begin{array}{c}
t_1 \\ t_2
\end{array} \right) \nonumber \;. \label{rotationtadpole}
\end{equation}
where $\cos\beta=v_1/v$ and $\sin\beta=v_2/v$. From Eqs.~(\ref{tadpole1})
and ~(\ref{tadpole2}) we have
\begin{mathletters}
\begin{eqnarray}
T_1&=&{1\over v}\left [ (m_{H_1}^2+\mu^2)v_1^2-(m_{H_2}^2+\mu^2)v_2^2
+{1\over 8}(g^2+g^{\prime 2})v^2(v_1^2-v_2^2) \right ]\nonumber \;, \\
&=&v\left [ (m_{H_1}^2+\mu^2)\cos ^2\beta-(m_{H_2}^2+\mu^2)\sin ^2\beta
+{1\over 8}(g^2+g^{\prime 2})v^2\cos 2\beta \right ]\;, \label{tadpoleM} \\
T_2&=&{1\over v}\left [ (m_{H_1}^2+m_{H_2}^2+2\mu ^2)v_1v_2+B\mu v^2
\right ]\nonumber \;, \\
&=&v\left [ {1\over 2}(m_{H_1}^2+m_{H_2}^2+2\mu ^2)\sin 2\beta+B\mu
\right ]\;.
\label{tadpoleMp}
\end{eqnarray}
\end{mathletters}
We see that the rotation (\ref{rotationtadpole}) through the angle
$\beta $ conveniently places all of the
dependence on gauge
couplings (D-terms) in $T_1$.
Setting $T_1=0$ and dividing by
$v\cos 2\beta$ yields the familiar tree-level condition
\begin{eqnarray}
{1\over 2}M_Z^2&=&{{m_{H_1}^2-m_{H_2}^2\tan ^2\beta }
\over {\tan ^2\beta -1}}-\mu ^2 \;. \label{treemin1}
\end{eqnarray}
Setting $T_2=0$ and dividing by
$v$ the other tree-level condition
\begin{eqnarray}
-B\mu &=&{1\over 2}(m_{H_1}^2+m_{H_2}^2+2\mu ^2)\sin 2\beta \;,
\label{treemin2}
\end{eqnarray}
is obtained. Notice that the signs of $B$ and $\mu $ are not determined by
the minimization conditions (only the relative sign is known),
giving rise to two distinct cases.

We can extend the above technique to include one-loop contributions to the
Higgs potential, deriving equations analogous to (\ref{treemin1})
and (\ref{treemin2}) by setting to zero linear combinations of
tadpoles rotated through
the angle $\beta$.
The one-loop effective potential is given by
\begin{eqnarray}
V_1&=&V_0+\Delta V_1\;,
\end{eqnarray}
where $V_0$ is the tree-level Higgs potential and
\begin{eqnarray}
\Delta V_1&=&{1\over {64\pi ^2}}{\rm Str}\left [{\cal M}^4
\left (\ln {{{\cal M}^2}\over {Q^2}}-{3\over 2}\right )\right ]\;, \label{V1}
\end{eqnarray}
is the one-loop contribution given in the dimensional reduction
($\overline{DR}$) renormalization scheme\cite{dimred}.
The supertrace is defined as
${\rm Str} f({\cal M}^2)=\sum _iC_i(-1)^{2s_i}(2s_i+1)f(m_i^2)$ where
$C_i$ is the color degrees of freedom and $s_i$ is the spin of the
$i^{th}$ particle.
To determine the minimum one must set the first derivatives of the
effective potential to zero
\begin{eqnarray}
{{\partial V_1}\over {\partial \psi }}=
{{\partial V_0}\over {\partial \psi }}+
{1\over {32\pi ^2}}{\rm Str}\left [
{{\partial {\cal M}^2}\over {\partial \psi }}
{\cal M}^2
\left (\ln {{{\cal M}^2}\over {Q^2}}-1\right )
\right ]=0\;. \label{V1d1}
\end{eqnarray}
We note that $f(m_i^2)$ usually involves the mass eigenstates of the
theory; one therefore ought to use the coupling of the Higgs fields to
the mass eigenstates in tadpole calculations in order to facilitate
comparisons between minimization techniques.
Evaluated at the minimum of $V_1$,
tadpole contributions involve the coupling $\partial {\cal M}^2/\partial\psi$
and the usual integration factor
${1\over {32\pi ^2}}{\cal M}^2\left (\ln {{{\cal M}^2}\over {Q^2}}-1\right )$;
setting tadpole contributions to zero is therefore equivalent to minimizing
the potential.
More generally, the $n$th derivatives of the effective potential
are related to the diagrams
(at zero external momentum) with $n$ external lines;
the minimization conditions at one-loop are
obtained by calculating diagrams with only one external line --
the tadpoles\cite{tadpole,sher}.

In order to maintain the linear combinations in (\ref{rotationtadpole}) for
the tree level relations, we
calculate with appropriate combinations of Higgs fields in
the external Higgs line in the tadpole diagrams. The Feynman rules
usually express these external Higgs lines as the physical Higgs
bosons $H$ or $h$, which are obtained from the Higgs fields $\psi_1,\psi_2$
by a rotation by an
angle $\alpha $ (in the opposite direction to the rotation $\beta$
performed above)
\begin{equation}
\left( \begin{array}{c}
H \\ h
\end{array} \right)= \left( \begin{array}{c@{\quad}c}
\cos \alpha & \sin \alpha  \\
-\sin \alpha  & \cos \alpha
\end{array} \right)\left( \begin{array}{c}
\psi_1 \\ \psi_2
\end{array} \right) \nonumber \;. \label{rotationhiggs}
\end{equation}
As with the tree-level tadpoles, we need to rotate the one-loop
contributions by
the same angle $\beta$ in order to express the minimization conditions most
simply. We therefore define the desired linear combinations ${\cal J}$,
${\cal J}_{\perp}$
of Higgs fields
\begin{equation}
\left( \begin{array}{c}
{\cal J} \\ {\cal J}_{\perp }
\end{array} \right)= \left( \begin{array}{c@{\quad}c}
\cos \beta  & -\sin \beta  \\
\sin \beta  & \cos \beta
\end{array} \right)\left( \begin{array}{c}
\psi_1 \\ \psi_2
\end{array} \right)= \left( \begin{array}{c@{\quad}c}
\cos (\beta + \alpha ) & -\sin (\beta + \alpha ) \\
\sin (\beta + \alpha ) & \cos (\beta + \alpha )
\end{array} \right)\left( \begin{array}{c}
H \\ h
\end{array} \right) \nonumber \;. \label{rotation1}
\end{equation}
To include the one-loop corrections,
we calculate the tadpole diagrams in Figure 2, and add the suitably
regularized result to the tree-level results.

\begin{center}
\epsfxsize=1in
\hspace*{0in}
\epsffile{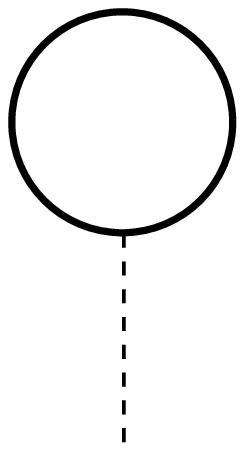}

\parbox{5.5in}{\small Fig.~2. The one-loop tadpole diagram. The loop consists
of matter and gauge-Higgs contributions.\\}
\end{center}

This tadpole technique is not new, and is equivalent to procedures followed
previously. However it provides an alternate
way of organizing the calculation and of understanding why the contributions
have their particular form. Moreover, the analytical
expressions obtained with the tadpole technique are often very useful,
particularly in certain regions of parameter space that are difficult
to explore by simply minimizing the potential numerically (e.g.
the low-$\tan\beta$ fixed-point region).

The method of determining the minimization conditions at one-loop
by calculating tadpoles is
especially convenient for including the corrections from the gauge and Higgs
sectors. The loop integrals are standard, and the only work is to determine the
coupling between the particle in the loop and the Higgs bosons ${\cal J}$ and
${\cal J}_{\perp}$ in Eq.~(\ref{rotation1}).
This approach is easier than including
the field dependent masses in the formal expression in Eq.~(\ref{V1}) and then
numerically finding the potential
minimum. On the other hand, calculating tadpoles
alone determines only the first derivatives of the one-loop Higgs potential,
and does not yield by itself the Higgs potential away from the minimum.
Fortunately, the minimization conditions are all one needs for many analyses.

It is crucial to include the one-loop corrections in the effective potential
in determining the vevs.
As shown by Gamberini, Ridolfi, and Zwirner\cite{grz}, the tree-level
Higgs vevs
$v_1$ and $v_2$ are very sensitive to the scale at which the renormalization
group equations are evaluated.
Thus it is necessary to determine the proper scale at which there are no large
logarithms so that the tree-level
results are reliable. As is well known, there is
a simple hierarchy of scales in these theories. As the soft-supersymmetry
breaking parameters are evolved down from the high scale, the Higgs potential
evolves so that an asymmetric minimum develops at some scale $\mu _0$. This
scale is determined by the condition
\begin{eqnarray}
m_{1}^2(\mu _0)m_{2}^2(\mu _0)-B^2\mu ^2(\mu _0)=0\;. \label{cond1}
\end{eqnarray}
For $Q>\mu _0$, the vevs $v_1$ and $v_2$ vanish.
For $Q<\mu _0$ the vevs
become nonzero. In the supergravity theories under consideration $m_{H_2}^2$
becomes negative, allowing Eq.~(\ref{cond1}) to be satisfied.
Figure 3 describes the potential in the regions of interest\cite{jorge}.

\begin{center}
\epsfxsize=5.75in
\hspace*{0in}
\epsffile{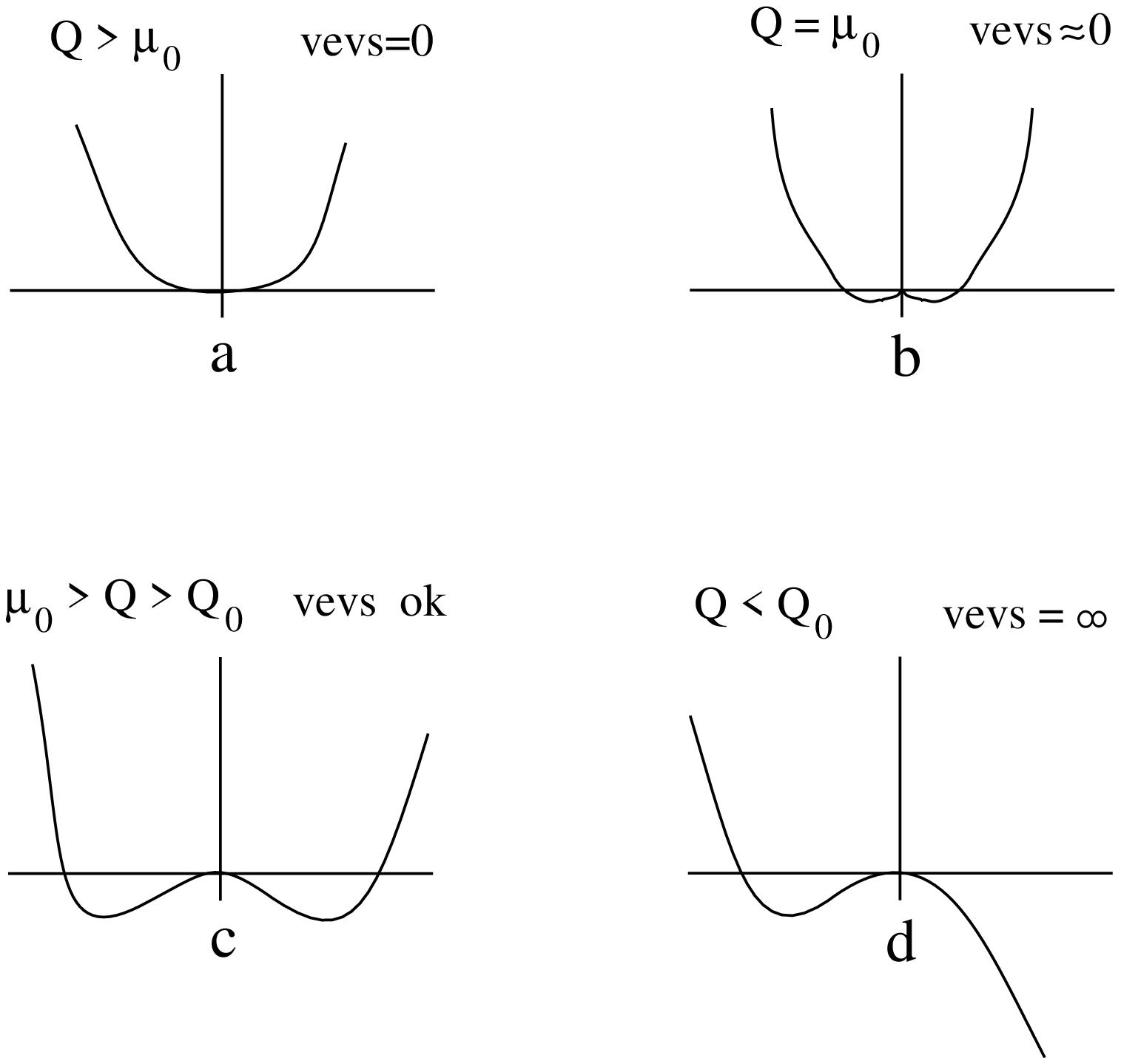}

\parbox{5.5in}{\small Fig.~3. (a) The vevs vanish for $Q > \mu _0$.
(b) For $Q\approx \mu _0$, the vevs become nonvanishing but small.
(c) For some scale $Q$ in the range $Q_0 < Q < \mu _0$ the vevs have the
correct magnitudes to give correct electroweak symmetry breaking.
(d) For $Q<Q_0$ the potential becomes unbounded from below.\\}
\end{center}

At some lower scale $Q_0<\mu _0$, the Higgs potential becomes unbounded from
below. The
scale $Q_0$ at which this occurs is determined by the condition
\begin{eqnarray}
m_{1}^2(Q_0)+m_{2}^2(Q_0)-2|B(Q_0)\mu (Q_0)|=0\;. \label{cond2}
\end{eqnarray}
This implies that in the tree-level potential
the vevs $v_1$ and $v_2$ must be driven off to
infinity because the potential becomes unbounded from below. Because the
vevs evolve from zero at or above
the scale $\mu _0$ all the way to infinity at
the scale $Q_0$, the vevs are very sensitive to the scale at which
they are evaluated.

The solution to this conundrum was provided in Ref.~\cite{grz}. The inclusion
of the one-loop contributions to the Higgs potential stabilizes the vevs
with respect to the scale $Q$ at which the parameters (which evolve according
to renormalization group equations) are evaluated.
The standard three cases considered are
(i) $M_{\rm SUSY}<Q_0<\mu _0$,
(ii) $Q_0<M_{\rm SUSY}<\mu _0$, and
(iii) $Q_0<\mu _0<M_{\rm SUSY}$.
In case (i) the scale $Q_0$ is determined by dimensional transmutation in the
sense of Coleman and Weinberg\cite{cw}. It was initially
realized that the one-loop contributions were important in this case,
because the
minimum of the Higgs potential is driven to the flat direction (``D-flat'') at
$\tan \beta =1$\cite{noscale2},
and it was crucial to include the one-loop contribution to lift this
degeneracy. This yields a light Higgs boson at tree-level (exactly zero
mass if $\tan \beta =1$), which is still acceptable
experimentally when the one-loop corrections to the Higgs boson mass are
included\cite{dh}. However the predicted SUSY mass spectrum is light and
already experimentally excluded.
Case (ii) has been the subject of much recent
work. Case (iii) is not of interest since electroweak symmetry
breaking does not occur.

To determine the minimum of the potential,
we include the one-loop tadpole contributions
\begin{mathletters}
\begin{eqnarray}
T_1+\Delta T_1&=&0\;, \\
T_2+\Delta T_2&=&0\;.
\end{eqnarray}
\end{mathletters}
The contributions $\Delta T_1$ and $\Delta T_2$ are given in the Appendix.

\section{Absence of fine-tuning}
The requirement that the supergravity model not be fine-tuned has been
recently
applied to limit the region of parameter space. This constraint requires that
the scale of supersymmetry breaking not be too high. Obtaining
reasonable criteria for declaring a particular theory unnaturally fine-tuned
remains a subject of debate.

The fine-tuning constraint becomes particularly restrictive in the small and
large $\tan \beta $ regions. For small $\tan \beta $ (near one), the Higgs
potential has its minimum near the D-flat direction. This implies
naturally large vevs. Then there must be a cancellation between
the two large terms on the right hand side of Eq.~(\ref{treemin1})
to obtain the
experimentally observed value for $M_Z$. Hence
for $\tan \beta \rightarrow 1$, the supersymmetric Higgs mass parameter $\mu $
must be tuned ever more precisely -- the fine-tuning problem. In this
section we discuss the various attempts to quantify this constraint.

The kinds of criteria
advocated by other authors are as follows
\begin{itemize}

\item Barbieri and Giudice\cite{bg} introduced a naturalness criteria
\begin{eqnarray}
\left | {{a_i}\over {M_Z^2}}{{\partial M_Z^2}\over
{\partial a_i}}\right |<\Delta &&\;,
\end{eqnarray}
for various fundamental parameters $a_i=m_0$, $m_{1\over 2}$, $\mu ^G$,
$A^G$, $B^G$ to obtain
an upper bound on the supersymmetric particle masses. They required that
$\Delta < 10$, i.e. no cancellations greater than an order of magnitude.

\item Lopez, et al.\cite{aspects} define several fine-tuning coefficients
e.g.
\begin{eqnarray}
{{\delta M_Z}\over {M_Z}}&=&c_{\mu }{{\delta \mu}\over {\mu }}\;. \label{ft1}
\end{eqnarray}
They show that a reasonable upper bound on
the simplest coefficient $c_{\mu }$ implies an upper bound
on $\mu $.

\item Arnowitt and Nath\cite{arnowittnath} require that $m_0^{}< 1$ TeV, a
condition that is easily applied phenomenologically.

\item Ross-Roberts\cite{RR} and de Carlos and Casas\cite{cc} consider
the fine-tuning of $M_Z$ in terms of $\lambda _t$
\begin{eqnarray}
{{\delta M_Z^2}\over {M_Z^2}}&=&c{{\delta \lambda _t^2}\over {\lambda _t^2}}\;,
\label{ft2}
\end{eqnarray}
where $c$ is required to be less than some small number e.g.
$c\alt 10$. Ross and Roberts who work strictly with the
tree-level Higgs potential argue that $\tan \beta \agt 2$, while
de Carlos and Casas argue that
the one-loop corrections to the Higgs potential ameliorate the fine-tuning.

\item Olechowski and Pokorski\cite{op} look at a full set of derivatives as in
Eq.~(\ref{ft1}), (\ref{ft2})
\begin{eqnarray}
{{\delta Q_j}\over {Q_j}}&=&\Delta _{ij}{{\delta P_i}\over {P_i}}\;,
\label{ftop}
\end{eqnarray}
where the $Q_j$ are the electroweak scale parameters $\lambda _t$,
$\lambda _{\tau} $, $v$, $\tan \beta $, $M_A^{}$, $M_Q^{}$, $M_U^{}$, and
the $P_i$ are the GUT scale parameters $\lambda _t^G$,
$\lambda _{\tau}^G $, $m_{1\over 2}$, $m_0^{}$, $\mu ^G$, $A^G$, $B^G$.
They also find that small $\tan \beta $ tends to be more unnatural, and
moreover for large values of $\tan \beta $, near where the top and bottom
quark couplings are equal, that the model becomes rapidly more fine-tuned as
$\tan \beta $ is increased.
These constraints are clearly quite involved. While they test a panoply of
fine-tuning relations, we feel they are overly complex for such a qualitative
and arbitrary notion as naturalness. Therefore we abandon this notion in
favor of a more intuitive definition similar to Lopez. et al.\cite{aspects}.

\item Casta\~{n}o, Piard, and Ramond\cite{Ramond}
choose a numerical definition in which the number of iterations
the computer has to find a solution is limited. It is not obvious how this
algorithm compares quantitatively to those defined above.
\end{itemize}

Our physical definition of naturalness
is simply $|\mu(M_Z)|\simeq |\mu(m_t)| < 500$ GeV.
A measure of the reasonableness of this
definition is the effect that small changes in $\mu$ have on $M_Z$. From the
tree level equation for $M_Z^{}$ (see Eqn.~(\ref{treemin1}))
it is readily
apparent that larger values of $|\mu| $ become more
unnatural.  In Figure 4 we plot the dependence of $M_Z^{}$ on $\mu $
and $B$ for both the tree-level calculation and the full one-loop
contributions to the Higgs potential. It can be seen that the one-loop
contributions reduce the fine-tuning to an extent.
This comparison
can be related to the plots of the vevs as a function of scale $Q$ as
discussed in Ref.~\cite{cc}.

\begin{center}
\epsfxsize=6in
\hspace*{0in}
\epsffile{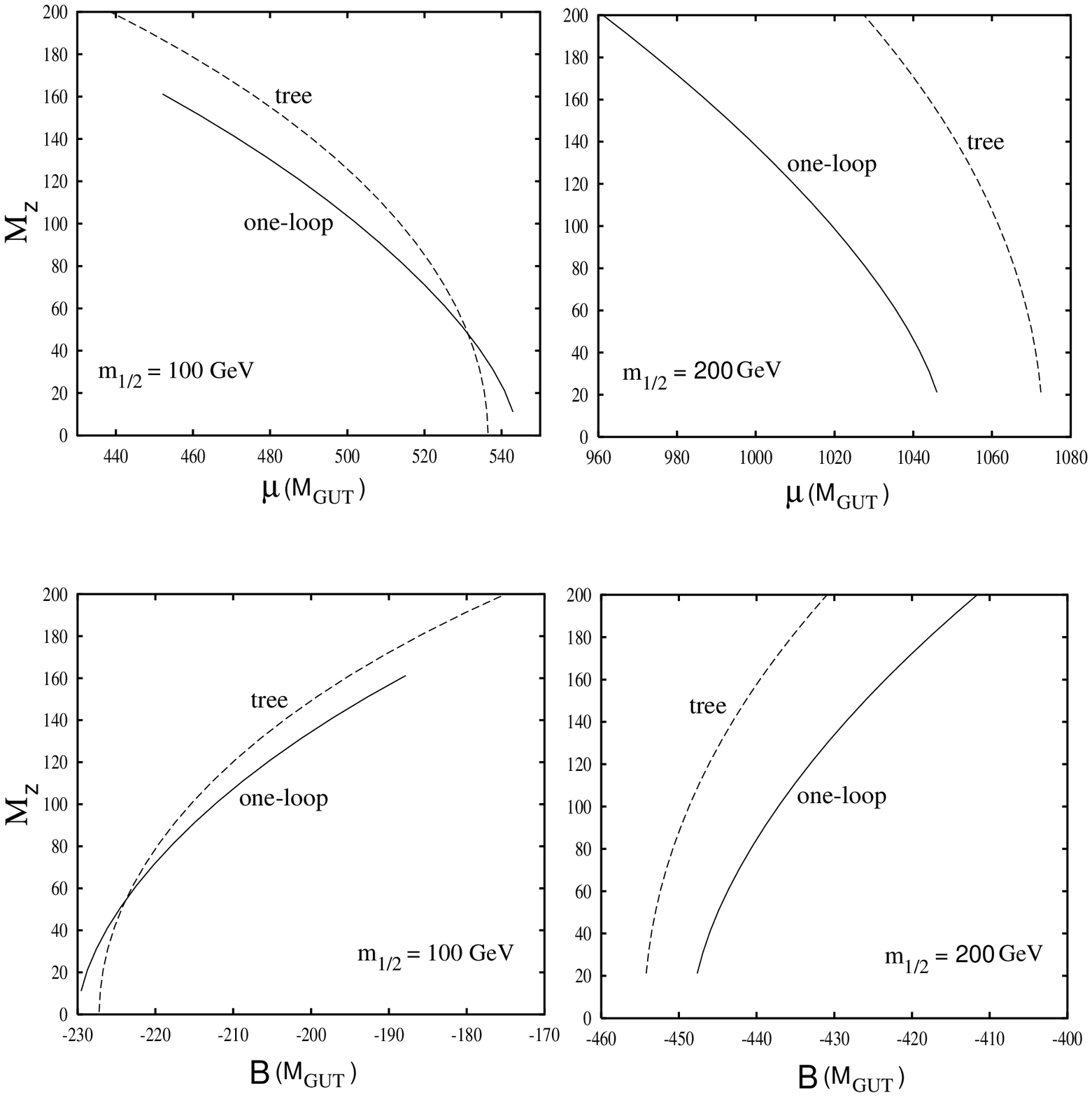}

\parbox{5.5in}{\small Fig.~4. The change in $M_Z$ with $\mu$ and $B$
for the fixed-point solution $m_t(m_t) = 160$, $\tan \beta = 1.47$,
$m_0 = 0$, $A^G = 0$, and
two different values of $m_{1 \over 2}$. The values of $\mu(M_G)$ meet our
naturalness criterion $|\mu(M_Z)| < 500$ GeV.
The solid (dashed) curves are the
results at one-loop (tree) level. The case $\mu < 0$ gives comparable
curves.\\}
\end{center}

The plots in Figure 4 correspond to a low-$\tan\beta$
fixed-point solution\cite{bbo}--\cite{BCPW};
in such cases the tree-level and one-loop-level  values turn out
to be comparable for either $\mu $ and $B$
in the region defined by
our naturalness criterion (only the degree of fine-tuning
changes). Consequently, including the one-loop corrections in the Higgs
potential does not have a critical
impact on the phenomenology.
This result does not extend to other regions of the
$m_t$-$\tan \beta$ plane since there our criterion for naturalness implies
a larger allowed range of $m_0$ and $m_{1 \over 2}$ in which the one-loop
contributions can change $\mu$ and $B$ significantly (see section 7).

One can consider quantitative
fine-tuning criteria analogous to those considered above,
\begin{eqnarray}
{{\delta M_Z}\over {M_Z}}&=&c_{\mu}^{}{{\delta \mu^G}\over {\mu^G}}\;,
\nonumber \\
{{\delta M_Z}\over {M_Z}}&=&c_B^{}{{\delta B^G}\over {B^G}}\;,
\label{ftus}
\end{eqnarray}
with e.g. $|c_{\mu }|$, $|c_B|^{}\alt 30$,
where the derivatives on the left hand side
are obtained at the physical $Z$ mass scale and the derivatives on the right
hand side are at the GUT scale (denoted by the $G$
superscript).
Since the RGE equation for $\mu$ is
proportional to $\mu$, the value of $\delta \mu/\mu$ is scale independent, but
$\delta B/B$ depends on scale. Table 1 gives
the values of $|c_{\mu}|$ and $|c_B|$
determined for the tree-level and one-loop curves for the low-$\tan \beta $
fixed point solution of Figure 4.
\vskip 0.5in
{\center \begin{tabular}{|c||c|c||c|c|}
\hline
\multicolumn{1}{|c|}{$m_{1 \over 2}$}
&\multicolumn{1}{|c|}{$|c_{\mu}|, (\mu >$ 0)}
&\multicolumn{1}{|c|}{$|c_B|, (\mu >$ 0)}
&\multicolumn{1}{|c|}{$|c_{\mu}|, (\mu <$ 0)}
&\multicolumn{1}{|c|}{$|c_B|, (\mu <$ 0)}
\\ \hline \hline
\multicolumn{1}{|c|}{100 GeV (loop)}
&\multicolumn{1}{|c|}{8.8}
&\multicolumn{1}{|c|}{8.2}
&\multicolumn{1}{|c|}{7.3}
&\multicolumn{1}{|c|}{5.2}
\\ \hline
\multicolumn{1}{|c|}{100 GeV (tree)}
&\multicolumn{1}{|c|}{13.5}
&\multicolumn{1}{|c|}{10.9}
&\multicolumn{1}{|c|}{13.5}
&\multicolumn{1}{|c|}{6.3}
\\ \hline \hline
\multicolumn{1}{|c|}{200 GeV (loop)}
&\multicolumn{1}{|c|}{25.6}
&\multicolumn{1}{|c|}{27.0}
&\multicolumn{1}{|c|}{20.3}
&\multicolumn{1}{|c|}{16.9}
\\ \hline
\multicolumn{1}{|c|}{200 GeV (tree)}
&\multicolumn{1}{|c|}{57.0}
&\multicolumn{1}{|c|}{46.8}
&\multicolumn{1}{|c|}{57.0}
&\multicolumn{1}{|c|}{27.6}
\\ \hline
\end{tabular}
\vskip .3in }
\begin{center}
{\bf Table 1:}  Values of $|c_{\mu}|$ and $|c_B|$ obtained at tree \\
and one-loop levels for the
fixed-point solution of Figure 5.
\end{center}
\vskip .5in
Note that inclusion of the full one-loop contribution substantially
reduces the fine-tuning constants $|c_{\mu}|$ and $|c_B|$.
Our entries for $|c_{\mu}|$ are somewhat larger than those found in
Ref.~\cite{aspects} because our model has a value of
$\tan \beta$ that is closer
to $\tan \beta =1$.

\section{Models}

The introduction of supersymmetry introduces many new unknown parameters
to the standard model. The advantage of the popular supergravity models is
that this number of new parameters is reduced to five or less.
The models discussed here should
only be viewed as examples of possible supersymmetry breaking scenarios.
Some features may be more general, however.

\noindent
{\bf A. General model\cite{radiative} }

The universal parameters at the GUT scale are
$m_0^{}$, $A^G$, $m_{1\over 2}$, $\mu ^G$, $B^G$.
In the minimal supergravity model, these five parameters describe
the higgsino and gaugino sectors. The universality of the scalar masses at the
GUT scale provides for the suppression of dangerous flavor changing neutral
currents involving the squarks of the first two generations.

\noindent
{\bf B. No-scale\cite{noscale2,noscale} }

In no-scale models two of the five parameters
are zero at the unification scale,
\begin{equation}
m_0=0,\quad A^G=0\;.
\end{equation}
Thus the scalar fields are massless there,
and $m_{1\over 2}$ is the sole origin of supersymmetry breaking.

\noindent
{\bf C. Strict no-scale\cite{noscale2,noscale} }

The strict no-scale model is a version of the no-scale model with
\begin{eqnarray}
B^G&=&0\:,
\end{eqnarray}
at the unification scale.

\noindent
{\bf D. Dilaton\cite{dilaton} }

When the dilaton $S$ receives a vev, one encounters a breaking of
supersymmetry that is of a different nature than that of
the minimal supergravity
scenarios described above.
The dilaton F-term scenario leads to simple boundary
conditions for the soft-supersymmetry parameters
\begin{equation}
m_0={1\over \sqrt{3}}m_{1\over 2},\quad A^G=-m_{1\over 2}\;.
\end{equation}
This model therefore has only three parameters.
When it is required that $\mu $ receive contributions
from supergravity only, the additional unification constraint
\begin{equation}
B^G=2m_0\;,
\end{equation}
is obtained. The dilaton version of supersymmetry breaking has
been studied in the MSSM in Ref.~\cite{blm} and for the flipped SU(5)
model in Ref.~\cite{lnz}.

\noindent
{\bf E. String-Inspired }

Supersymmetry breaking in strings is a nonperturbative effect, since
supersymmetry is preserved order by order in
perturbation theory. Very little is known about nonperturbative effects in
string theory. Recently the authors of Ref.~\cite{bim} have proposed to
parametrize our ignorance of the exact nature of the breakdown of
supersymmetry.
The dilaton breaking scenario
above is a specific case of more general scenario
of supersymmetry breaking in which the moduli fields $T_m$ also receive a vev.
If one restricts oneself to the case where only one $T$ field
and the dilaton $S$
get vevs, then
the amount of SUSY breaking that arises from each sector can be parametrized
by the ``goldstino angle\cite{bim}'' $\theta $. The dilaton breaking case
corresponds to $\sin \theta =1$.
The angle $\theta $ is constrained by low-energy phenomenology since
purely dilaton breaking gives a universal boundary condition for the
scalar masses, and the breaking of supersymmetry when the moduli field gets a
vev will gives
rise to FCNCs in the low-energy theory. According to Ref.~\cite{bim}
the more general case, where substantial contributions to supersymmetry
breaking arise from the moduli field getting a vev, is not
ruled out.

The unification scale in the string-inspired model is roughly an
order of magnitude higher than the scale at which the gauge couplings unify
in the MSSM. Presumably large threshold corrections due to non-degenerate GUT
particles could account for this
discrepancy.

\noindent
{\bf F. Large $\tan \beta $ scenario\cite{tbt} }

The correct electroweak symmetry breaking does not
occur for too large values
of $\tan \beta $. If $\tan \beta \agt m_t(m_t)/m_b(m_t)$,
then the bottom quark Yukawa drives the Higgs masses parameter $m_1^2$ negative
first (instead of $m_2^2$ from the top quark Yukawa coupling).
For $\tan \beta $ close to this limit considerable fine-tuning is
required to get the correct electroweak scale. This situation is ameliorated
somewhat with the inclusion of the one-loop corrections in the effective
potential\cite{op}.

\section{Ambidextrous Approach to RGE Integration}

Previous RGE studies of the supersymmetric particle spectrum have evolved from
inputs at the GUT scale (the top-down method\cite{aspects,Ramond})
or from inputs at the electroweak
scale (the bottom-up approach\cite{op}).
Our approach incorporates some boundary conditions at both electroweak and
GUT scales, which we call the ambidextrous approach.
We specify
$m_t$ and $\tan \beta$ at the electroweak scale (along with $M_Z$ and $M_W$)
and $m_{1 \over 2}$,
$m_0$, and $A^G$ at the GUT scale.
The soft supersymmetry breaking parameters are evolved from the GUT scale to
the electroweak scale and then $\mu (M_Z)$ and $B(M_Z)$
(or $\mu (m_t)$ and $B (m_t)$) are determined by
the tadpole equations at one-loop order.
Subsequently $\mu $ and $B$ can be RGE-evolved
up to the GUT scale. This strategy is effective because the RGEs for the
soft-supersymmetry breaking parameters (see the appendix) do not depend on
$\mu $ and $B$. This method has two powerful advantages:
First, any point in the $m_t$ -- $\tan \beta$ plane can be readily
investigated in specific supergravity
models since $m_t$ and $\tan \beta $ are taken as inputs.
Second, the tadpole equations Eq.~(\ref{t1p}-\ref{t2p})
are easy to solve in the ambidextrous approach.
The $T_1$ equation can be solved iteratively for $\mu(M_Z)$, and
then the $T_2$ equation explicitly gives $B(M_Z)$.
We stress the numerical simplicity: no derivatives need be calculated
and no functions need to be numerically minimized.

We now describe our numerical approach in more detail.
Starting with our low-energy choices for $m_t$, $\tan \beta$, $\alpha_3$,
and $m_b$ (and using the experimentally determined values for
$\alpha_1$, $\alpha_2$ and $m_\tau$\cite{pdb}), we integrate
the MSSM RGEs from $m_t$ to $M_G$ with $M_G^{}$ taken
to be the scale $Q$ at which  $\alpha_1 (Q)$ =  $\alpha_2 (Q)$.
We then specify $m_{1 \over 2}$, $m_0$, and $A^G$, and integrate
back down to $m_t$ where we solve the tadpole equations for $\mu(m_t)$
and $B(m_t)$. We can then integrate the RGEs back to $M_G$
to obtain  $\mu^G$ and $B^G$ at $M_G^{}$. A few remarks are pertinent:

1) We integrate the two-loop MSSM RGEs for the gauge and Yukawa
couplings\cite{bbo}, but only the one-loop MSSM RGEs (as given in the appendix)
for the other
supersymmetric parameters. We retain the important two-loop gauge
and Yukawa effects (only the two-loop gaugino RGEs exist\cite{yamada}
and we desire to be consistent with regard to the order for the
soft-supersymmetry breaking parameters).

2) Since we are working only to one-loop order in most RGEs, we neglect
threshold effects at both the GUT\cite{barhall}--\cite{Hagiwara}
and supersymmetry scales. To properly
take into account threshold effects requires that the appropriate
beta functions be calculated at every (s)particle threshold, and since
the complete supersymmetric mass spectrum and the
GUT particle spectra are  generally not known a priori,
the calculation and use of these beta functions is extremely daunting.
In any event, it should not be critical to incorporate these
threshold effects into a one-loop calculation since two-loop
contributions will be comparable to threshold effects.

3) We take the lower bound of our integration at $m_t$ instead of $M_Z$
for several reasons. As shown by several
groups\cite{grz,aspects,op,Ramond,cc}, inclusion
of the one-loop effects into the effective potential makes electroweak
symmetry breaking roughly independent of scale; the scale $Q=m_t$ is roughly
the value for which the large logs cancel among themselves in the one-loop
corrections to the minimization conditions.
We choose $m_t$ as the boundary since the RGEs (in particular for the
gauge and Yukawa couplings) are simple at scales above $m_t$, and it is
non-trivial to extend them below $m_t$. In addition, the
choice of $m_t$ facilitates comparison with previous work on gauge and Yukawa
unification and fixed points.

\section{Results}

We discuss the supersymmetric spectrum and phenomenology for several
representative points in the $m_t$ -- $\tan\beta$ plane. For the most
part we focus on the low-$\tan \beta$ fixed-point region since
it is very attractive to explain the large top quark mass as a fixed
point phenomenon\cite{bbo}--\cite{BCPW}. Moreover, the supersymmetric
spectrum in this
region is largely unexplored, probably due to fears of excessive
fine-tuning. However, as addressed in Section 4, these
fears are not necessarily justified; there remains substantial
viable parameter space for which fine-tuning does not pose great concern,
particularly with the inclusion of the full one-loop corrections to
the effective potential\cite{op}.

The $m_t$ -- $\tan \beta$ parameter space can be divided into
several distinct regions, as shown in Figure 5.

\begin{center}
\epsfxsize=6in
\hspace*{0in}
\epsffile{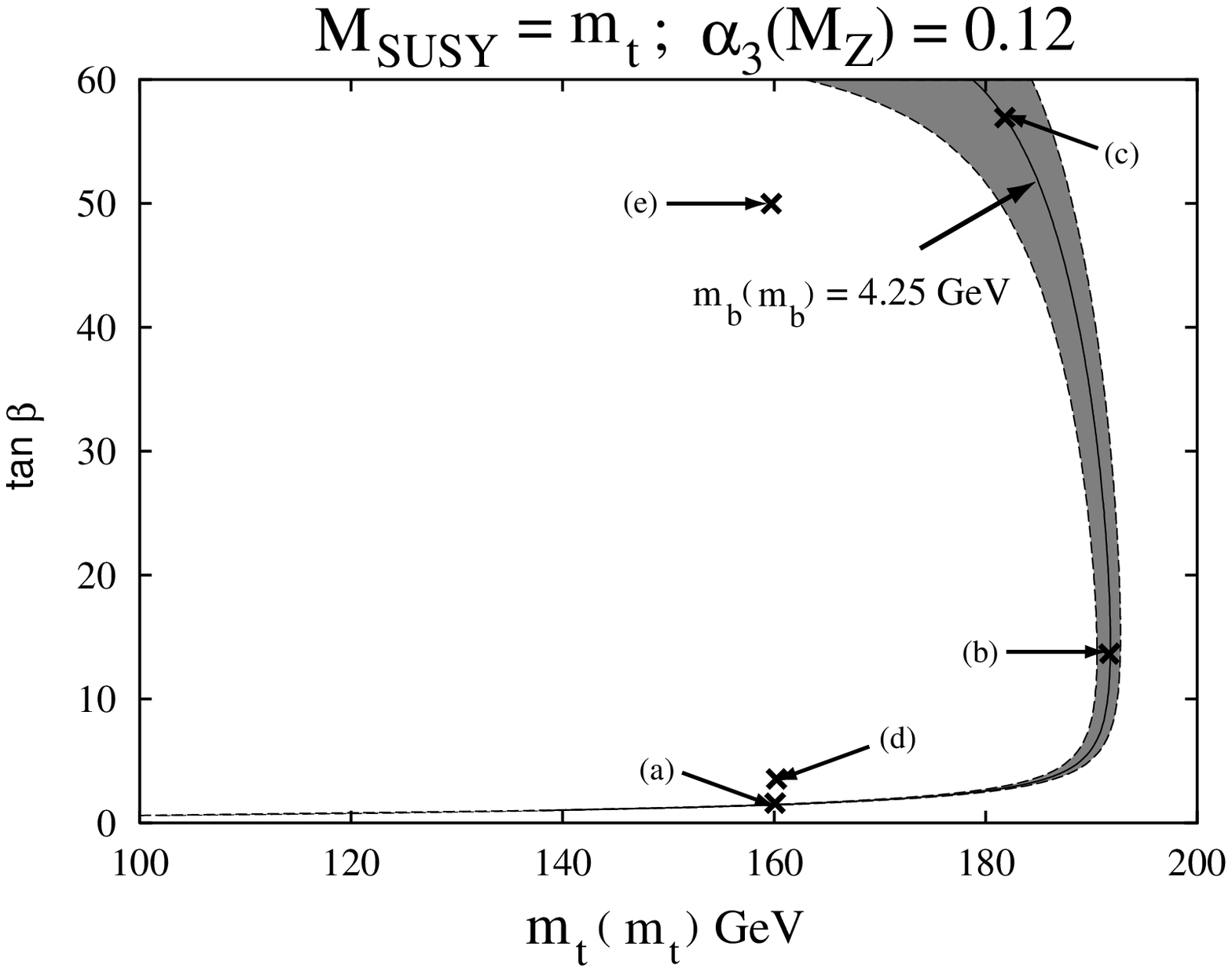}

\parbox{5.5in}{\small Fig.~5. The allowed $m_t$ -- $\tan \beta$ parameter
space assuming Yukawa unification $\lambda_b(M_G)$ = $\lambda_\tau(M_G)$.
\cite{bbo}. The shaded area indicates the region for which
$m_b(m_b) = 4.25\pm 0.15$ GeV. Points representative of distinct regions
within this parameter space are denoted with labels (a)-(e).\\}
\end{center}

We discuss
the supersymmetric mass spectrum for each of these regions.
Unless otherwise specified, we take $A^G = 0$, $\alpha_3(M_Z^{}) = 0.120$,
and $m_b(m_b) = 4.25$ GeV. The qualitative behaviour in each region
should not depend greatly on these parameters.

\begin{itemize}

\item {\bf (a) Low-$\tan \beta$ Fixed Point}

As a typical example of the low $\tan \beta$ fixed values region we consider
the point $m_t(m_t) = 160$ GeV, $\tan \beta = 1.47$ (for which
$\lambda_t(M_G) = 2.7$).
We aim to determine the GUT-scale parameter space for
which this solution can be obtained from the minimization of the
effective potential. Using the tadpole method, we explore a grid
of $m_0$ and $m_{1 \over 2}$ values and apply both experimental and
naturalness bounds. For the lower experimental limits, we adopt
the values listed in Table 2 following Ref.~\cite{borz}.
\vskip 0.5in
{\center \begin{tabular}{|c|c|}
\hline
\multicolumn{1}{|c|}{Particle}
&\multicolumn{1}{|c|}{Experimental Limit (GeV)}
\\ \hline \hline
\multicolumn{1}{|c|}{gluino}
&\multicolumn{1}{|c|}{120}
\\ \hline
\multicolumn{1}{|c|}{squark, slepton}
&\multicolumn{1}{|c|}{45}
\\ \hline
\multicolumn{1}{|c|}{chargino}
&\multicolumn{1}{|c|}{45}
\\ \hline
\multicolumn{1}{|c|}{neutralino}
&\multicolumn{1}{|c|}{20}
\\ \hline
\multicolumn{1}{|c|}{light higgs}
&\multicolumn{1}{|c|}{60}
\\ \hline
\end{tabular}
\vskip .3in }
\begin{center}
{\bf Table 2:}  Approximate experimental bounds that we apply in Figure 6.
\end{center}
\vskip .5in
Figure 6 shows the allowed parameter space for both signs of $\mu$
along with the most restrictive constraints in each case.
The contours of constant $|\mu |$ are ellipses in the $m_0-m_{1\over 2}$
plane for $|\mu |>>M_Z$.

\begin{center}
\epsfxsize=5.1in
\hspace*{0in}
\epsffile{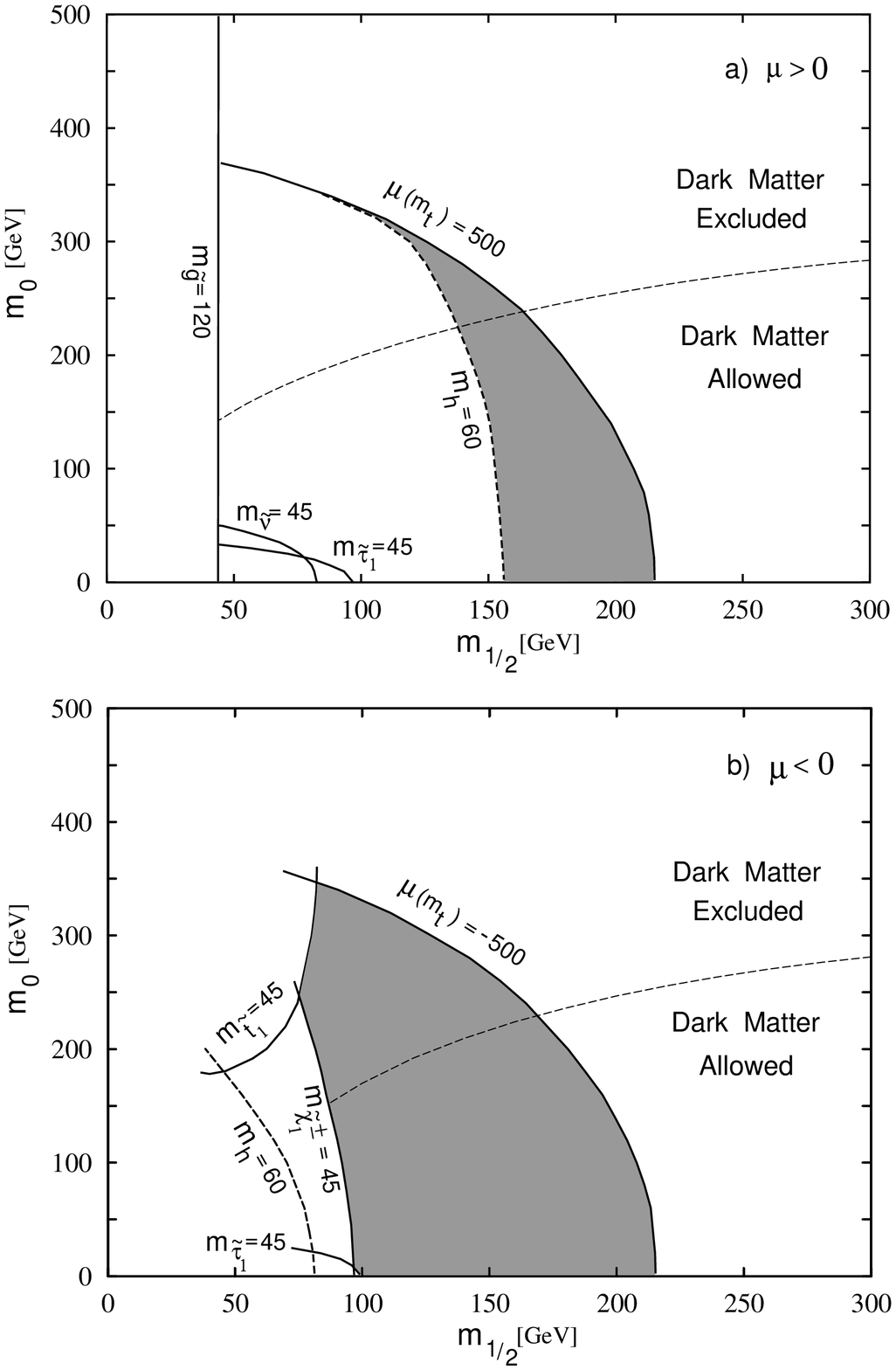}

\parbox{5.5in}{\small Fig.~6. The allowed $m_0$, $m_{1 \over 2}$
region is shaded for the low-$\tan\beta$ fixed point
$m_t(m_t) = 160$ GeV, $\tan \beta = 1.47$ solution with (a)
$\mu > 0$ and (b) $\mu < 0$. The experimental
bounds in Table 2 and the naturalness bound $|\mu (m_t)|<500$ GeV are imposed
with $A^G = 0$ GeV.
A semi-quantitative dark matter constraint (given by
Eq.~(\ref{manuel})) is also shown.}
\end{center}

Note that the $\mu < 0$ case has more available parameter space;
it is also slightly more $\it{natural}$, as indicated from
the fine-tuning constants given in Table 1.

\hspace*{0.3in} We have indicated the light
scalar higgs experimental limit with
a dashed line in Figure 6; care must be taken when enforcing this particular
constraint since the allowed parameter space is
somewhat sensitive to the exact $m_h^{}$ limit.
Moreover, the $m_h^{}$ bound includes only the one-loop
quark-squark contributions
given in reference\cite{berz},
and it is expected that inclusion of the chargino and neutralino
contributions can affect the mass of the
light scalar higgs by a few  GeV\cite{HH}. In
general an increase in $m_h$ (as in all the experimental
constraints) tends to increasingly restrict the allowed parameter space,
while a decrease in $m_h$ has the opposite effect.

\hspace*{0.3in} Overall we find
substantial phenomenologically viable parameter space, especially for
$\mu < 0$.
However the maximal values of the GUT parameters $m_0$ and $m_{1\over 2}$
are not large ($m_0\alt 350$ GeV, $m_{1\over 2} \alt 225$ GeV) implying
a rather
light low energy supersymmetric mass spectrum.
Also included in Figure 6 is the semi-quantitative dark matter constraint
of Drees and Nojiri \cite{dn} (see Eq.~(\ref{manuel}) below).
For this low $\tan\beta$ fixed-point case it implies that
$m_0 \alt$ 250 GeV, though this approximate
bound ought not to be taken strictly.

\hspace*{0.3in} We now investigate
the supersymmetric particle mass spectra
dependence on $m_0$ and $m_{1 \over 2}$
independently for this low $\tan\beta$ fixed-point solution.
Figure 7 shows the dependence of the
supersymmetric spectrum on $m_{1 \over 2}$ in the noscale model
($m_0 = 0$). For the squarks and sleptons we plot the lightest mass
eigenstates; in addition we plot the heaviest stop,
$m_{\tilde{t}_2}$, for reference. We label the
chargino and neutralino masses such that
$M_{\chi_1^{\pm}}<M_{\chi_2^{\pm}}$ and
$M_{\chi_1^0}<M_{\chi_2^0}<M_{\chi_3^0}<M_{\chi_4^0}$.

\begin{center}
\epsfxsize=5.6in
\hspace*{0in}
\epsffile{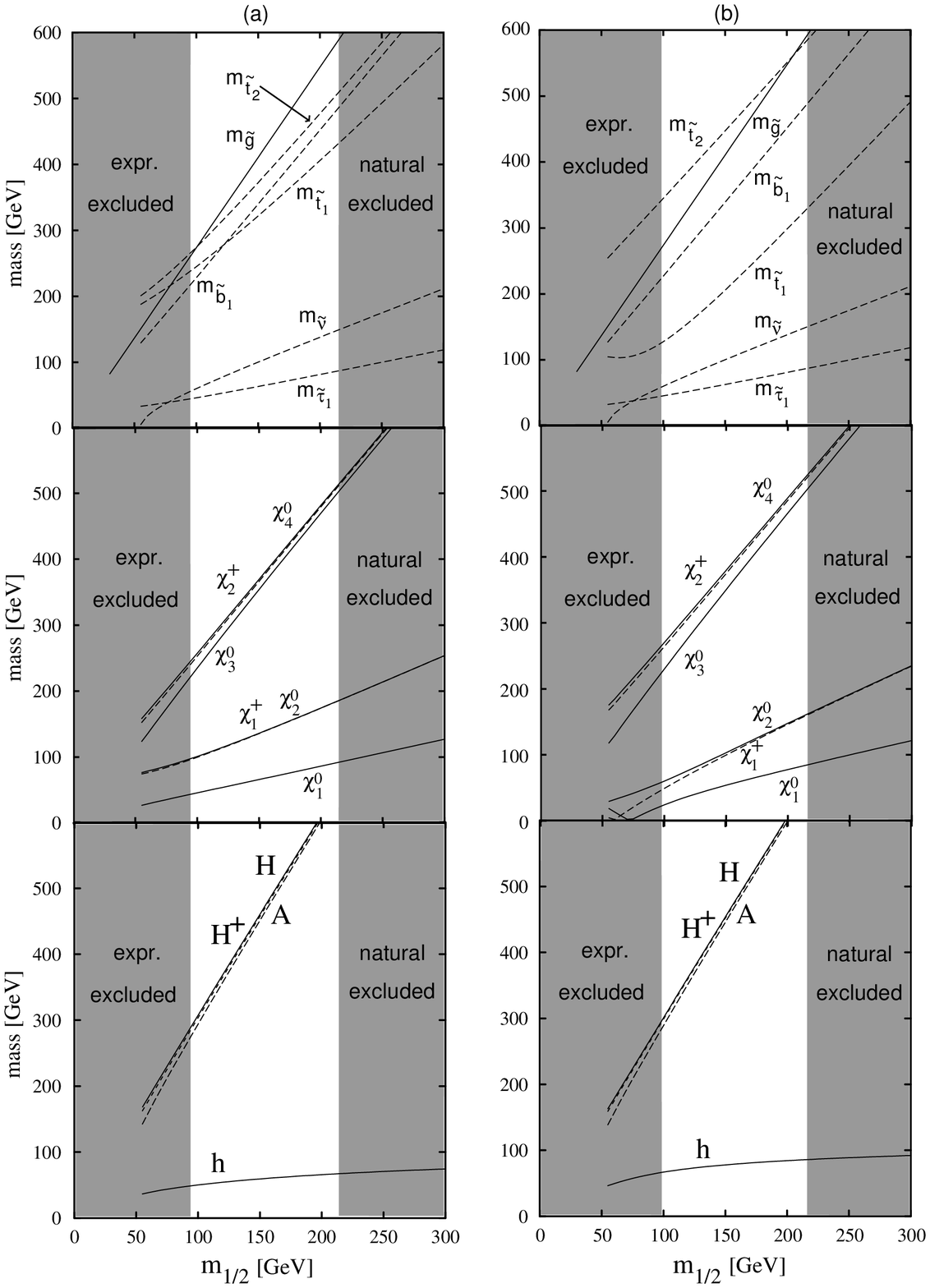}

\parbox{5.5in}{\small Fig.~7. The low $\tan\beta$ fixed point
solutions for (a)
$\mu > 0$ and (b) $\mu < 0$ with $m_0 = 0$ GeV and $A^G = 0$.
The experimentally excluded region includes all experimental constraints
except for the bound on $m_h$, since it is sensitive to chargino and
neutralino contributions\cite{HH}.}
\end{center}

\hspace*{0.3in} Figure 8 shows all the
squark and slepton masses for the same low-$\tan\beta$
fixed-point solution with $\mu >$ 0. Note that the squarks of the first two
generations can be heavier than those of the third;
the up and charm squarks are degenerate as are the
down and strange squarks.
The slepton masses
are approximately generation independent in this case, though this
need not be true in general (e.g. see Table 3 below).

\begin{center}
\epsfxsize=6in
\hspace*{0in}
\epsffile{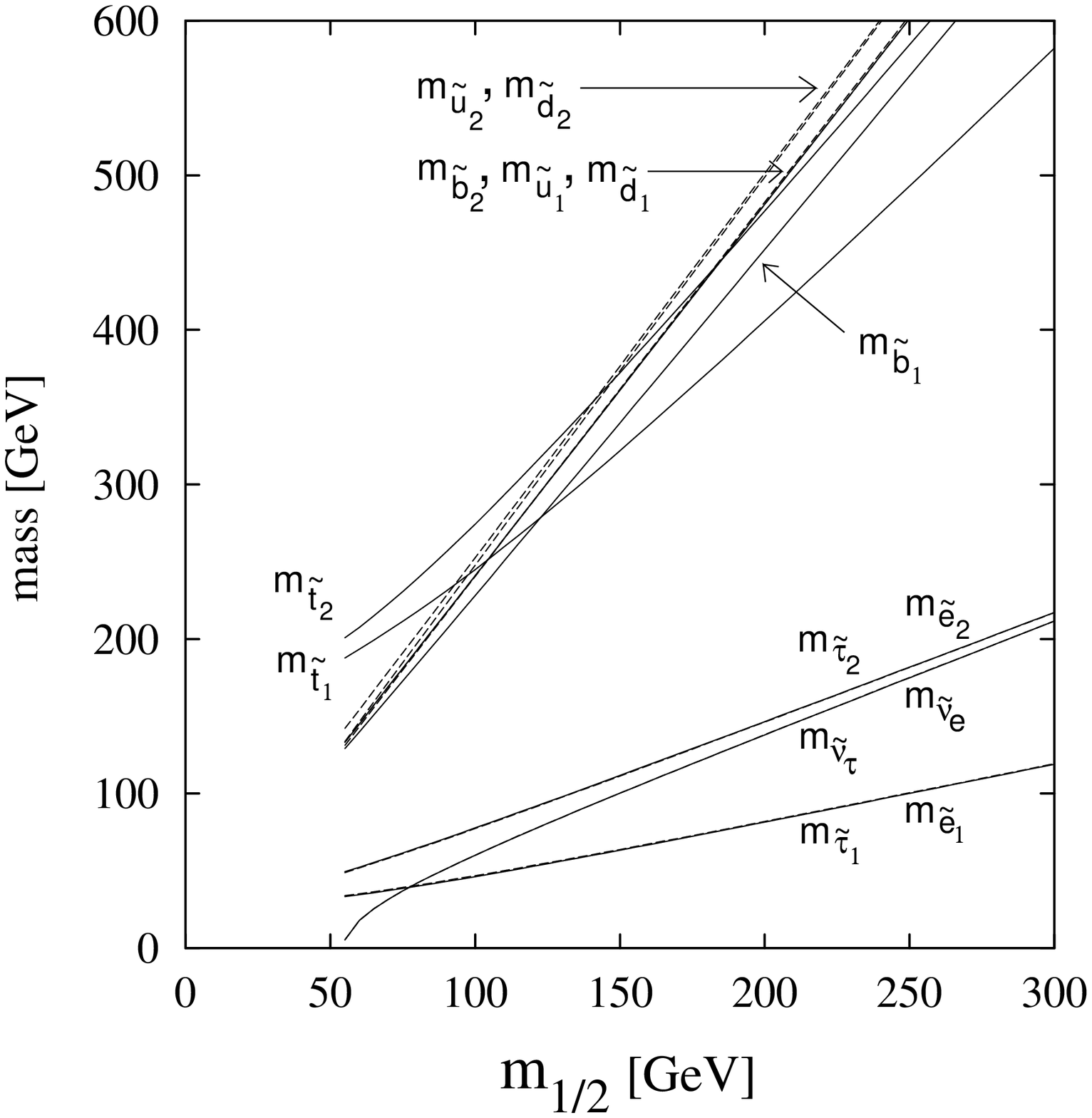}

\parbox{5.5in}{\small Fig.~8. The squark and slepton masses for the
low-$\tan\beta$ fixed-point solution in the no-scale model with $\mu > 0$.
The solid (dashed) lines correspond to the third (first and second)
generation.
The excluded regions are the same as in the previous figure.}
\end{center}

\hspace*{0.3in} Figure 9 illustrates the dependence of the
supersymmetric spectrum on $m_0$.
(Here we take $m_{1 \over 2} = 150$ GeV.)
The mass of most of the SUSY particles increase with increasing $m_0$ (see,
e.g. Eq.~(\ref{fi})).

\begin{center}
\epsfxsize=5.6in
\hspace*{0in}
\epsffile{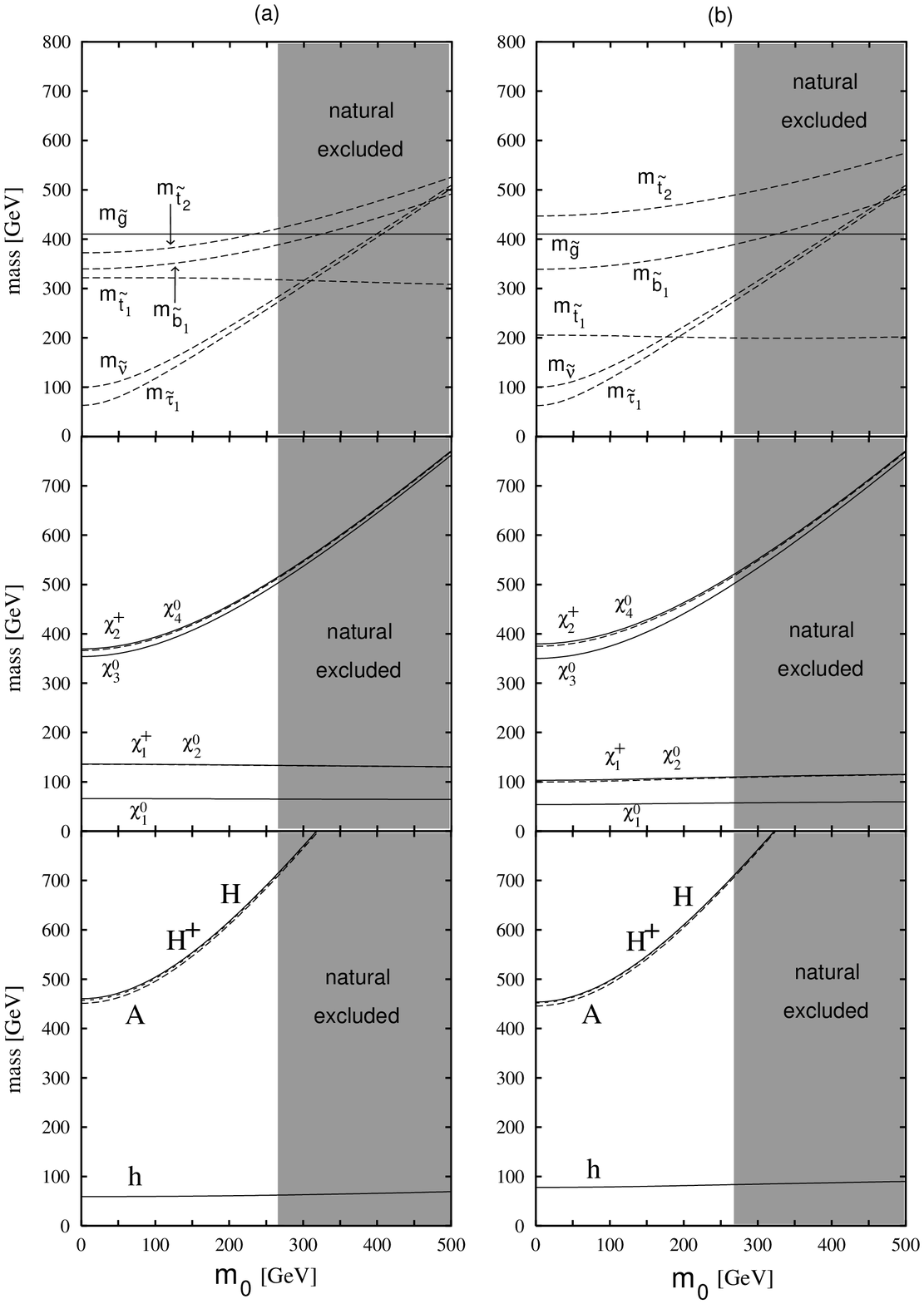}

\parbox{5.5in}{\small Fig.~9. The low $\tan \beta$
fixed point solutions for (a)
$\mu > 0$ and (b) $\mu < 0$ with $m_{1 \over 2} = 150$ GeV and $A^G = 0$.
The shaded area denotes the region excluded by our naturalness criterion.}
\end{center}

\hspace*{0.3in} We also give qualitative
descriptions of the allowed parameter space in the
other significant $m_t$ -- $\tan \beta$ regions.

\item {\bf (b) Medium-$\tan \beta$ Fixed Point}

The allowed $m_0$ -- $m_{1 \over 2}$ parameter space is substantially larger
in this case than it is for the low-$\tan\beta$ fixed point.
Our naturalness condition
allows substantially larger values of $m_0$ and $m_{1 \over 2}$
($m_0\alt 725$ GeV and $m_{1 \over 2}\alt 325$ GeV for $m_t(m_t) = $ 192 GeV,
and
$\tan\beta = $ 15); however, dark matter constraints will still
require $m_0 \alt$ 300 GeV.
For $\mu >$ 0, experimental bounds on $m_{\tilde{\nu}}$, $m_{\tilde{\tau}_1}$,
and
$m_{\chi_{1^{\pm}}}$ push the lower bound for $m_0$ and $m_{1 \over 2}$
up slightly.
For $\mu < 0$, experimental bounds for $m_{\tilde{\nu}}$ and
$m_{\tilde{\tau}_1}$ also become more restrictive, but the $m_{\chi_{1^{\pm}}}$
and $m_{\tilde{t}_1}$ constraints
become less restrictive.
In both cases the constraint from the lightest scalar higgs
mass becomes less restrictive; even in the
$\mu > 0$ case, it will not play an important role in limiting the
allowed $m_0$ and $m_{1 \over 2}$ region.
To summarize, the medium-$\tan\beta$ fixed point region allows
larger values of $m_0$ and $m_{1 \over 2}$ from our naturalness
constraint while the experimental restrictions on this parameter
space do not change much from the low-$\tan\beta$ case.

\item {\bf (c) High-$\tan \beta$ Fixed Point}

This region describes the SO(10) relation $\lambda_t$ =
$\lambda_b$ = $\lambda_\tau$ where $\lambda_i$ $\agt$ 1.
There is not much parameter space
remaining without weakening our naturalness condition.
For the case $m_t(m_t) = 178$ GeV, $\tan\beta = 61$
(with $m_0=400$ GeV, $m_{1 \over 2}=400$ GeV,
and $\mu(m_t) \approx 575$ GeV)
the particle spectrum is given in the following table.
\vskip 0.5in
{\center \begin{tabular}{|c|c|}
\hline
\multicolumn{1}{|c|}{Particle}
&\multicolumn{1}{|c|}{Mass (GeV)}
\\ \hline \hline
\multicolumn{1}{|c|}{gluino}
&\multicolumn{1}{|c|}{1078}
\\ \hline
\multicolumn{1}{|c|}{stop, sbottom}
&\multicolumn{1}{|c|}{751,900; 763,881}
\\ \hline
\multicolumn{1}{|c|}{up squarks, down squarks}
&\multicolumn{1}{|c|}{1029,1060; 1026,1063}
\\ \hline
\multicolumn{1}{|c|}{stau, tau sneutrino}
&\multicolumn{1}{|c|}{183,454; 417}
\\ \hline
\multicolumn{1}{|c|}{other sleptons}
&\multicolumn{1}{|c|}{431,494; 487}
\\ \hline
\multicolumn{1}{|c|}{charginos}
&\multicolumn{1}{|c|}{323,590}
\\ \hline
\multicolumn{1}{|c|}{neutralinos}
&\multicolumn{1}{|c|}{167,323,579,588}
\\ \hline
\multicolumn{1}{|c|}{higgs: $m_A$,$m_{H^{\pm}}$,$m_H$,$m_h$}
&\multicolumn{1}{|c|}{364,377,363,131}
\\ \hline
\end{tabular}
\vskip .3in }
\begin{center}
{\bf Table 3:}  Particle spectrum for $m_t =$ 178 GeV, $\tan\beta = 61$  \\
(where $m_0 = 400$ GeV, $m_{1 \over 2} = 400$ GeV, $A^G$ = 0).
\end{center}
\vskip .5in
As before the above particle spectrum is calculated at the
scale $m_t$.
We obtain no natural solutions for $\mu < 0$.

\item {\bf (d) Low-$\tan \beta$, not a Fixed Point}

This region has a large amount of viable parameter
space; naturalness bounds allow substantially higher values of
$m_{1 \over 2}$ ($m_{1 \over 2}\alt 330$ GeV
for $m_t(m_t) = 160$ GeV with $\tan\beta = $ 3), and experimental
constraints do not
further restrict this parameter space to any great extent
(though the sneutrino and chargino bounds are pushed upward somewhat).
In fact, the higgs constraint is weakened a great deal for
$\mu > 0$, allowing relatively low values for
$m_0$ and $m_{1 \over 2}$. Moreover,
the light stop constraint is less important in the $\mu < 0$
case.

\item {\bf (e) High-$\tan \beta$, not a Fixed Point}

The allowed parameter space is reduced by the lightest
stau constraint (which cannot be the LSP),
though some parameter region remains.
The allowed parameter space is bounded by chargino, stau,
dark matter, and our naturalness
constraint which give
$180 \alt m_0 \alt 300$ GeV and $85 \alt m_{1 \over 2} \alt 400$ GeV
for $m_t(m_t) = 160$ GeV, $\tan\beta = 45$.
The light higgs and the light stop constraints are not important for
either sign of $\mu$.

\end{itemize}

In addition, we varied $A^G$ from $-500$ to $+500$ GeV
in the low $\tan \beta$ fixed point case;
we found little change in the resulting parameter space except
that the light stop constraint is more (less) restrictive for
$A^G$ negative (positive) and $\mu < 0$. The fixed point solution in
radiative electroweak symmetry breaking has also been studied recently
in Ref.~\cite{copw}.

A critical constraint\cite{borz,bbmr,bsg}
on the supersymmetric spectrum is the rare decay
$b\rightarrow s\gamma$. We remark here that
regions of the parameter space illustrated in the previous
figures are not ruled out by this constraint. This will be the
subject of a forthcoming paper\cite{bbop2}.

\section{SUSY Mass Spectrum Correlations}

For smaller values of $\tan \beta $ it is clear from the tree-level expression
Eq.~(\ref{treemin1}) that $|\mu|$ is usually large compared to the the
electroweak scale $M_Z$. Furthermore the fine-tuning problem
in this situation is softened when the one-loop contributions to the
Higgs potential are included. For values of $\mu $ just a few times larger
than $M_Z$, the particle spectrum is governed by certain asymptotic
behaviors which we discuss in this section.

As discussed previously, the gaugino masses are related (through one-loop
order) by the same ratios that describe the gauge couplings at the
electroweak scale. This observation, together with the fact that $|\mu|$ is
large, yields simple correlations between the lightest chargino and neutralinos
and the gluino\cite{arnowittnath,lnp}, namely
\begin{mathletters}
\begin{eqnarray}
M_{\chi _1^{0}}&\simeq &M_1 \;, \\
M_{\chi _1^{\pm}}&\simeq &M_{\chi _2^0}\;\;\simeq \;\;M_2\;\;=\;\;
{{\alpha _2}\over {\alpha_1}}M_1\;\simeq
\;\;2M_1\;\;\simeq \;\;M_{\chi _1^0} \;, \\
m_{\tilde{g}}&=&M_3\;\;=\;\;{{\alpha _3}\over {\alpha_2}}M_2\;\;=\;\;
{{\alpha _3}\over {\alpha_1}}M_1\;,
\end{eqnarray}
\end{mathletters}
where the quantities in these equations are evaluated at
scale $m_t$.
The heaviest chargino and the two heaviest neutralino states are primarily
Higgsino with
\begin{eqnarray}
M_{\chi _2^{\pm}}&\simeq &M_{\chi _3^0}\;\;\simeq \;\;M_{\chi _4^0}\;\;\simeq
\;\;|\mu | \;. \\
\end{eqnarray}

The lightest Higgs $h$ has small
mass for
$\tan \beta $ near one at tree-level by virtue of the D-flat direction;
its mass comes from
radiative corrections\cite{dh,lnpwz}. The heavy Higgs states are
(approximately) degenerate $\approx M_A$ because at tree-level
$M_A=-{{B\mu }\over {\sin 2\beta}}\approx -B\mu$ is large.

The squark and slepton masses also display
simple asymptotic behavior at large $|\mu|$.
The first and second squark generations are approximately degenerate
(though not degenerate enough
to ignore their contributions to the minimization of the effective potential).
The squark and slepton mass spectra are shown in Figures 7
through 9.
The splitting of the stop quark masses grows as $|\mu|$ increases.
This splitting of the sbottom states does not change much with $\mu $ for
small $\tan\beta$.

\section{Dark Matter}

The neutralino as the LSP is an ideal candidate for the dark matter since it
is stable and interacts weakly.
The MSSM utilizes R-parity conservation so the lightest neutralino
must annihilate to ordinary matter
($\chi \chi \rightarrow {\rm R-even}$ matter)
to a sufficient extent to avoid overclosing the universe\cite{dm}.
For a bino-like LSP,
dark matter considerations put an upper bound on the parameter $m_0$.
We adopt the conservative viewpoint that the
contribution of the LSP alone to the dark matter
of the universe must be less than the closure density.
Roberts and Roszkowski\cite{RoRo} apply an additional constraint in which
the neutralino is required to make up a substantial fraction
of the dark matter; this requirement
provides a lower bound on $m_0$ as well.
The recent results from COBE
suggest that the dark matter is a mixture of hot and cold dark matter.
Although it may be simpler to assume that all of the cold dark matter is
composed of one contribution, it is perhaps premature to assume this.
We remark that the recent exciting claims of
experimental evidence for dark matter in nearby galaxies\cite{MACHO,EROS}
only solves the
local baryonic dark matter problem\cite{Turner}. The origin of the
nonbaryonic dark matter needed to close the universe is still unknown.

The typical situation in the low $\tan \beta$ fixed point solutions is
that $|\mu |>> M_2$ and consequently the lightest neutralino
(which is the LSP) is predominantly
gaugino; indeed the LSP is predominantly bino.
The neutralino mass matrix is
\begin{equation}
M_N=\left( \begin{array}{c@{\quad}c@{\quad}c@{\quad}c}
M_1 & 0 & -M_Z\cos \beta \sin \theta _W^{} & M_Z\sin \beta \sin \theta _W^{}
\\
0 & M_2 & M_Z\cos \beta \cos \theta _W^{} & -M_Z\sin \beta \cos \theta _W^{}
\\
-M_Z\cos \beta \sin \theta _W^{} & M_Z\cos \beta \cos \theta _W^{} & 0 & \mu
\\
M_Z\sin \beta \sin \theta _W^{} & -M_Z\sin \beta \cos \theta _W^{} & \mu & 0
\end{array} \right)\;. \nonumber \\
\end{equation}
For $|\mu |>> M_2$ the lightest two neutralinos are
predominantly bino and wino, and hence the bino and gaugino purities are
high.
In this case any mass limit on the
lightest neutralino from $Z$ decays at LEP disappears since the $Z$
couples only to the Higgsino component of the neutralino.
Figure 10 gives the bino and gaugino purities for the low-$\tan\beta$ fixed
point solution in the no-scale model, corresponding to Fig. 7.

\begin{center}
\epsfxsize=3.9in
\hspace*{0in}
\epsffile{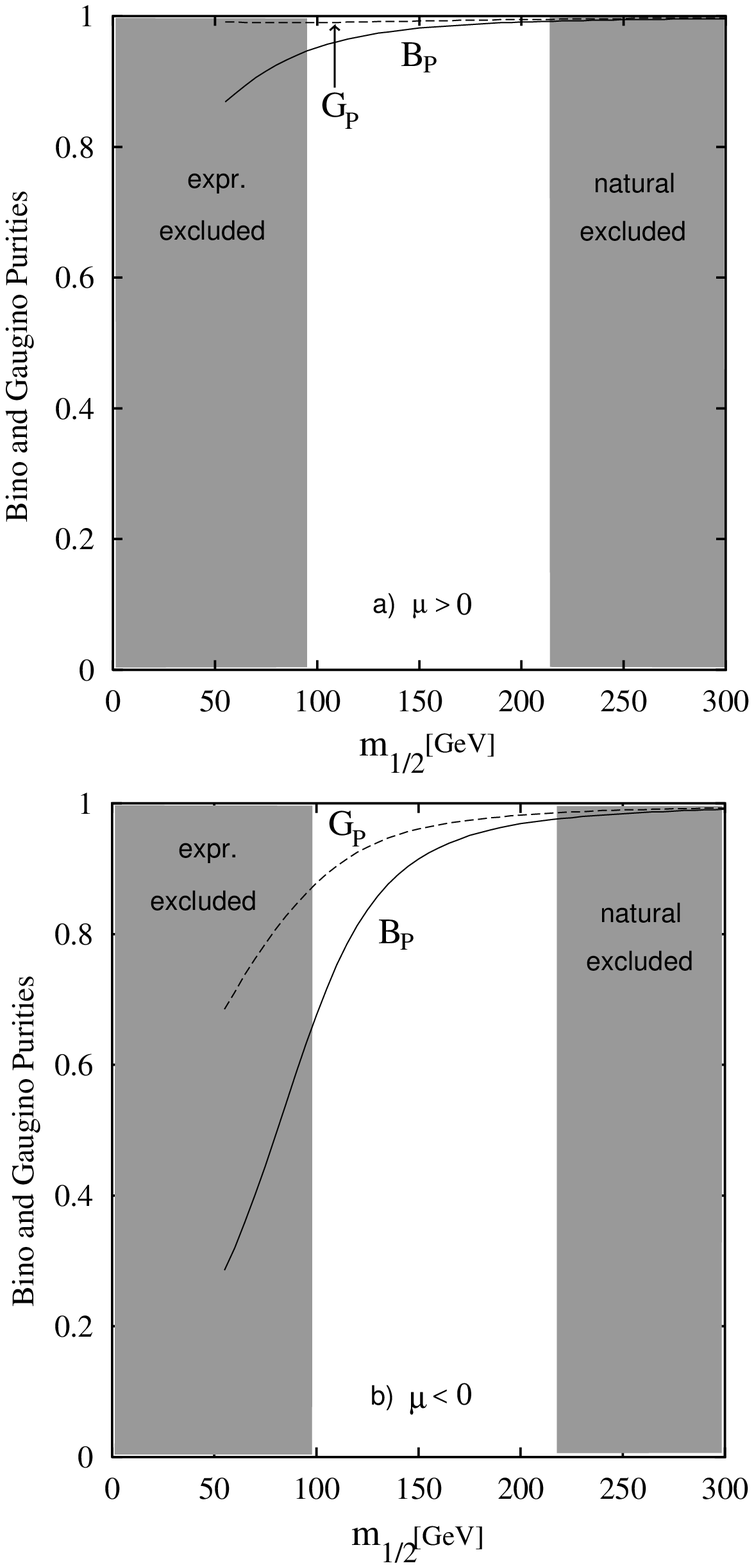}

\parbox{5.5in}{\small Fig.~10. The bino and gaugino purities for
the low $\tan\beta$ fixed point solution with $m_t = 160$ GeV,
$\tan\beta$ = 1.47 in the no-scale
model with (a) $\mu > 0$ and (b) $\mu < 0$.
Shaded regions are forbidden by experimental and fine-tuning considerations.}
\end{center}

Given that the solutions are comfortably in the high bino purity region, we
apply the semi-quantitative constraint of Drees and Nojiri\cite{dn}
(valid roughly for $|\mu |>m_{1\over 2}$, $M_{\chi _1^0}>60$ GeV)
\begin{eqnarray}
&&{{(m_0^2+1.83M_{\chi _1^0}^2)^2}\over {M_{\chi _1^0}^2\left [
\left (1-{{M_{\chi _1^0}^2}\over {m_0^2+1.83M_{\chi _1^0}^2}}\right )^2+
\left ({{M_{\chi _1^0}^2}\over {m_0^2+1.83M_{\chi _1^0}^2}}\right )^2
\right ]}}<1\times 10^6\;{\rm GeV}^2\;,  \label{manuel}
\end{eqnarray}
to obtain the line corresponding to $\Omega h^2=1$ in Figure 6.
This formula is based on the observation that for the bino-like LSP the
annihilation rate is dominated by the sleptons, and it
neglects a possible enhanced annihilation rate that may occur if there
are significant $s$-channel pole contributions.
The bino and gaugino purities for non-zero $m_0$ (in particular the dilaton
model) are similar to the above figures.

\section{Conclusion}
The motivation of this work has been to distill
the interesting
supersymmetric phenomenology of the low-$\tan\beta$ fixed-point
region that can explain the origin of a large top quark mass.
The RGEs are solved with some boundary conditions taken from both GUT
and low energy scales.
The minimization conditions
on the effective potential are obtained with the tadpole method.

Our principle findings can be summarized as follows:

\begin{itemize}

\item Solutions with a $\lambda _t$ fixed point,
$m_t \alt 170$ GeV and radiative breaking of the electroweak symmetry breaking
are allowed. These solutions are characterized by
relatively large values of the
supersymmetric Higgs mass parameter $|\mu|$, which implies that the
supersymmetric particle spectrum displays a simple asymptotic behavior.
The solutions also meet the naturalness criterion $|\mu(M_Z)| < 500$ GeV
for both signs of $\mu$.

\item Representative sparticle mass spectra are presented for the $\lambda _t$
fixed point solutions.

\item Over most of the GUT parameter space for the low-$\tan\beta$
fixed-point, the gaugino masses exhibit
show simple correlations due to the relatively large value $|\mu|$ compared
to $M_2$.
The heaviest chargino and the two heaviest neutralinos
have masses  approximately $|\mu|$; the lightest chargino and the second
lightest neutralino have masses approximately $M_2$; the lightest
neutralino (LSP) has a mass approximately $M_1\simeq M_2/2$.
The lightest Higgs
obtains its mass almost entirely from radiative corrections, and the
states $H$, $H^{\pm }$, $A$ are relatively heavy and approximately degenerate.

\item In the early universe
the LSP will annihilate sufficiently neglecting $s$-channel pole annihilation
for most of the parameter space
($m_0 \alt 300$ GeV) so as not to
overclose the universe.

\item The values of $\mu$ and $B$
derived from the one-loop Higgs
potential analyses are very similar to the tree-level results
in the low-$\tan\beta$ fixed-point region
when the parameters $M_Z$ and
$\tan \beta $ are taken as input.
However, the one-loop corrections to the Higgs potential somewhat
ameliorate the fine-tuning problem.

\item The tadpole method is a convenient way to calculate the
one-loop minimization conditions. We have obtained these conditions in an
analytic form including all contributions
from the gauge-Higgs sector and
matter multiplets.

\end{itemize}

\section{Acknowledgements}
We thank Manuel Drees for a discussion on dark matter and Roger Phillips
for collaboration on $b\rightarrow s\gamma$ analyses.
PO would like to thank Chris Kolda for discussions regarding
supersymmetric model building.
This research was supported
in part by the University of Wisconsin Research Committee with
funds granted by
the Wisconsin Alumni Research Foundation, in part by the U.S.~Department of
Energy under contract no.~DE-AC02-76ER00881, and in part by the Texas National
Laboratory Research Commission under grant nos.~RGFY93-221 and FCFY9302.
MSB was supported in
part by an SSC Fellowship.
PO was supported in
part by an NSF Graduate Fellowship.

\newpage
\baselineskip28pt

\section{Appendix}

\noindent {\bf Renormalization Group Equations}

The renormalization equations for the gauge couplings and the Yukawa couplings
to two-loop order can be found in Ref.~\cite{bbo}.
In the most general case the evolution equations involve matrices. For example
the Yukawa couplings form three-by-three Yukawa matrices: ${\bf U}$ for the
up-type quarks, ${\bf D}$ for the
down-type quarks, and ${\bf E}$ for the
charged leptons. Similarly the soft-supersymmetry breaking parameters form the
matrices ${\bf M}_{Q_L}$, ${\bf M}_{U_R}$, ${\bf M}_{D_R}$,
${\bf M}_{L_L}$, and ${\bf M}_{E_R}$. Finally there are in general
matrices for the trilinear soft-supersymmetry breaking ``A-terms'':
${\bf A_U}$, ${\bf A_D}$, and ${\bf A_E}$. It turns out to be useful
to define the combinations
${\bf U_A}_{ij}\equiv {\bf A_U}_{ij}{\bf U}_{ij}$, etc.
in the matrix version of the RGE's.
Then the  evolution of the soft-supersymmetry parameters (with our convention
for signs) is given by the
following
renormalization group equations\cite{bbmr}
\begin{mathletters}
\begin{eqnarray}
{{dM_i}\over {dt}}&=&{2\over {16\pi ^2}}b_ig_i^2M_i\;, \\
{{d{\bf U_A}}\over {dt}}&=&{1 \over {16\pi ^2}}
\Bigg [-\Big ({13\over 15}g_1^2+3g_2^2+{16\over 3}g_3^2\Big ){\bf U_A}
+2\Big ({13\over 15}g_1^2{M_1}+3g_2^2{M_2}
+{16\over 3}g_3^2{M_3}\Big ){\bf U}\nonumber \\
&&+\Big \{\Big [4({\bf U_AU^{\dagger }U})
+6{\bf Tr}({\bf U_AU^{\dagger }}){\bf U}\Big ]
+\Big [5({\bf UU^{\dagger }U_A})+3{\bf Tr}({\bf UU^{\dagger }}){\bf U_A}\Big ]
\nonumber \\
&&+2({\bf D_AD^{\dagger }U})+({\bf DD^{\dagger }U_A})\Big \}\Bigg ]\;, \\
{{d{\bf D_A}}\over {dt}}&=&{1 \over {16\pi ^2}}
\Bigg [-\Big ({7\over 15}g_1^2+3g_2^2+{16\over 3}g_3^2\Big ){\bf D_A}
+2\Big ({7\over 15}g_1^2{M_1}+3g_2^2{M_2}
+{16\over 3}g_3^2{M_3}\Big ){\bf D}\nonumber \\
&&+\Big \{\Big [4({\bf D_AD^{\dagger }D})
+6{\bf Tr}({\bf D_AD^{\dagger }}){\bf D}\Big ]
+\Big [5({\bf DD^{\dagger }D_A})+3{\bf Tr}({\bf DD^{\dagger }}){\bf D_A}\Big ]
\nonumber \\
&&+2({\bf U_AU^{\dagger }D})+({\bf UU^{\dagger }D_A})
+2{\bf Tr}({\bf E_AE^{\dagger }}){\bf D}+{\bf Tr}({\bf EE^{\dagger }})
{\bf D_A}
\Big \}\Bigg ]\;, \\
{{d{\bf E_A}}\over {dt}}&=&{1 \over {16\pi ^2}}
\Bigg [-\Big (3g_1^2+3g_2^2\Big ){\bf E_A}
+2\Big (3g_1^2{M_1}+3g_2^2{M_2}
\Big ){\bf E}\nonumber \\
&&+\Big \{\Big [4({\bf E_AE^{\dagger }E})
+2{\bf Tr}({\bf E_AE^{\dagger }}){\bf E}\Big ]
+\Big [5({\bf EE^{\dagger }E_A})+{\bf Tr}({\bf EE^{\dagger }}){\bf E_A}\Big ]
\nonumber \\
&&+6({\bf D_AD^{\dagger }E})+3({\bf DD^{\dagger }E_A})\Big \}\Bigg ]\;, \\
{{dB}\over {dt}}&=&{2\over {16\pi ^2}}\Big ({3\over 5}g_1^2M_1+3g_2^2M_2
+{\bf Tr}(3{\bf UU_A}+3{\bf DD_A}+{\bf EE_A})\Big )\;, \\
{{d\mu }\over {dt}}&=&{\mu \over {16\pi ^2}}\Big (-{3\over 5}g_1^2-3g_2^2
+{\bf Tr}(3{\bf UU^{\dagger }}+3{\bf DD^{\dagger }}
+{\bf EE^{\dagger }})\Big )\;, \\
{{dM_{H_1}^2}\over {dt}}&=&{2 \over {16\pi ^2}}
\Big (-{3\over 5}g_1^2M_1^2-3g_2^2M_2^2\nonumber \\
&&+3{\bf Tr}({\bf D}({\bf M}_{Q_L}^2+{\bf M}_{D_R}^2){\bf D^{\dagger}}
+M_{H_1}^2{\bf DD^{\dagger }}+{\bf D_AD_A^{\dagger }})\nonumber \\
&&+{\bf Tr}({\bf E}({\bf M}_{L_L}^2+{\bf M}_{E_R}^2){\bf E^{\dagger}}
+M_{H_1}^2{\bf EE^{\dagger }}+{\bf E_AE_A^{\dagger }})\Big )\;, \\
{{dM_{H_2}^2}\over {dt}}&=&{2 \over {16\pi ^2}}
\Big (-{3\over 5}g_1^2M_1^2-3g_2^2M_2^2\nonumber \\
&&+3{\bf Tr}({\bf U}({\bf M}_{Q_L}^2+{\bf M}_{U_R}^2){\bf U^{\dagger}}
+M_{H_2}^2{\bf UU^{\dagger }}+{\bf U_AU_A^{\dagger }})\Big )\;, \\
{{d{\bf M}_{Q_L}^2}\over {dt}}&=&{2 \over {16\pi ^2}}
\Big (-{1\over 15}g_1^2M_1^2-3g_2^2M_2^2-{16\over 3}g_3^2M_3^2\nonumber \\
&&+{1\over 2}[{\bf UU^{\dagger }}{\bf M}_{Q_L}^2+{\bf M}_{Q_L}^2
{\bf UU^{\dagger }}+2({\bf U}{\bf M}_{U_R}^2{\bf U^{\dagger }}
+m_{H_2}^2{\bf UU^{\dagger }}+{\bf U_A^{}U_A^{\dagger }})]\nonumber \\
&&+{1\over 2}[{\bf DD^{\dagger }}{\bf M}_{Q_L}^2+{\bf M}_{Q_L}^2
{\bf DD^{\dagger }}+2({\bf D}{\bf M}_{D_R}^2{\bf D^{\dagger }}
+m_2^2{\bf DD^{\dagger }}+{\bf D_A^{}D_A^{\dagger }})]\Big )\;, \\
{{d{\bf M}_{U_R}^2}\over {dt}}&=&{2 \over {16\pi ^2}}
\Big (-{16\over 15}g_1^2M_1^2-{16\over 3}g_3^2M_3^2\nonumber \\
&&+[{\bf U^{\dagger }U}{\bf M}_{U_R}^2+{\bf M}_{U_R}^2
{\bf U^{\dagger }U}+2({\bf U^{\dagger }}{\bf M}_{Q_L}^2{\bf U}
+m_{H_2}^2{\bf U^{\dagger }U}+{\bf U_A^{\dagger }U_A^{}})]\Big )\;, \\
{{d{\bf M}_{D_R}^2}\over {dt}}&=&{2 \over {16\pi ^2}}
\Big (-{4\over 15}g_1^2M_1^2-{16\over 3}g_3^2M_3^2\nonumber \\
&&+[{\bf D^{\dagger }D}{\bf M}_{D_R}^2+{\bf M}_{D_R}^2
{\bf D^{\dagger }D}+2({\bf D^{\dagger }}{\bf M}_{Q_L}^2{\bf D}
+m_{H_1}^2{\bf D^{\dagger }D}+{\bf D_A^{\dagger }D_A^{}})]\Big )\;, \\
{{d{\bf M}_{L_L}^2}\over {dt}}&=&{2 \over {16\pi ^2}}
\Big (-{3\over 5}g_1^2M_1^2-3g_2^2M_2^2\nonumber \\
&&+{1\over 2}[{\bf EE^{\dagger }}{\bf M}_{L_L}^2+{\bf M}_{L_L}^2
{\bf EE^{\dagger }}+2({\bf E}{\bf M}_{E_R}^2{\bf E^{\dagger }}
+m_{H_1}^2{\bf EE^{\dagger }}+{\bf E_A^{}E_A^{\dagger }})]\Big )\;, \\
{{d{\bf M}_{E_R}^2}\over {dt}}&=&{2 \over {16\pi ^2}}
\Big (-{12\over 5}g_1^2M_1^2\nonumber \\
&&+[{\bf E^{\dagger }E}{\bf M}_{E_R}^2+{\bf M}_{E_R}^2
{\bf E^{\dagger }E}+2({\bf E^{\dagger }}{\bf M}_{L_L}^2{\bf E}
+m_{H_1}^2{\bf E^{\dagger }E}+{\bf E_A^{\dagger }E_A^{}})]\Big )\;,
\end{eqnarray}
\end{mathletters}
For our purposes it is sufficient to consider these equations keeping only
the leading terms in the mass hierarchy in the three generation MSSM.
The resulting
renormalization group equations\cite{rge} are given below to leading order.
\begin{mathletters}
\begin{eqnarray}
{{dM_i}\over {dt}}&=&{2\over {16\pi ^2}}b_ig_i^2M_i\;, \\
{{dA_t}\over {dt}}&=&{2\over {16\pi ^2}}\Big (\sum c_ig_i^2M_i
+6\lambda _t^2A_t+\lambda _b^2A_b\Big )\;, \\
{{dA_b}\over {dt}}&=&{2\over {16\pi ^2}}\Big (\sum c_i^{\prime }g_i^2M_i
+6\lambda _b^2A_b+\lambda _t^2A_t+\lambda _{\tau}^2A_{\tau}\Big )\;, \\
{{dA_{\tau }}\over {dt}}&=&{2\over {16\pi ^2}}\Big (\sum c_i^{\prime \prime }
g_i^2M_i+3\lambda _b^2A_b+4\lambda _{\tau}^2A_{\tau}\Big )\;, \\
{{dB}\over {dt}}&=&{2\over {16\pi ^2}}\Big ({3\over 5}g_1^2M_1+3g_2^2M_2
+3\lambda _b^2A_b+3\lambda _t^2A_t+\lambda _{\tau}^2A_{\tau}\Big )\;, \\
{{d\mu }\over {dt}}&=&{\mu \over {16\pi ^2}}\Big (-{3\over 5}g_1^2-3g_2^2
+3\lambda _t^2+3\lambda _b^2+\lambda _{\tau}^2\Big )\;, \\
{{dM_{H_1}^2}\over {dt}}&=&{2 \over {16\pi ^2}}
\Big (-{3\over 5}g_1^2M_1^2-3g_2^2M_2^2
+3\lambda _b^2X_b+\lambda _{\tau}^2X_{\tau }\Big )\;, \\
{{dM_{H_2}^2}\over {dt}}&=&{2 \over {16\pi ^2}}
\Big (-{3\over 5}g_1^2M_1^2-3g_2^2M_2^2
+3\lambda _t^2X_t\Big )\;, \\
{{dM_{Q_L}^2}\over {dt}}&=&{2 \over {16\pi ^2}}
\Big (-{1\over 15}g_1^2M_1^2-3g_2^2M_2^2-{16\over 3}g_3^2M_3^2
+\lambda _t^2X_t+\lambda _b^2X_b\Big )\;, \\
{{dM_{t_R}^2}\over {dt}}&=&{2 \over {16\pi ^2}}
\Big (-{16\over 15}g_1^2M_1^2-{16\over 3}g_3^2M_3^2
+2\lambda _t^2X_t\Big )\;, \\
{{dM_{b_R}^2}\over {dt}}&=&{2 \over {16\pi ^2}}
\Big (-{4\over 15}g_1^2M_1^2-{16\over 3}g_3^2M_3^2
+2\lambda _b^2X_b\Big )\;, \\
{{dM_{L_L}^2}\over {dt}}&=&{2 \over {16\pi ^2}}
\Big (-{3\over 5}g_1^2M_1^2-3g_2^2M_2^2
+\lambda _{\tau}^2X_{\tau }\Big )\;, \\
{{dM_{\tau _R}^2}\over {dt}}&=&{2 \over {16\pi ^2}}
\Big (-{12\over 5}g_1^2M_1^2
+2\lambda _{\tau}^2X_{\tau }\Big )\;,
\end{eqnarray}
\end{mathletters}
and for the two light generations,
\begin{mathletters}
\begin{eqnarray}
{{dA_u}\over {dt}}&=&{2\over {16\pi ^2}}\Big (\sum c_ig_i^2M_i
+\lambda _t^2A_t\Big )\;, \\
{{dA_d}\over {dt}}&=&{2\over {16\pi ^2}}\Big (\sum c_i^{\prime }g_i^2M_i
+\lambda _b^2A_b+{1\over 3}\lambda _{\tau}^2A_{\tau}\Big )\;, \\
{{dA_e}\over {dt}}&=&{2\over {16\pi ^2}}\Big (\sum c_i^{\prime \prime }
g_i^2M_i+\lambda _b^2A_b+{1\over 3}\lambda _{\tau}^2A_{\tau}\Big )\;, \\
{{dM_{q_L}^2}\over {dt}}&=&{2 \over {16\pi ^2}}
\Big (-{1\over 15}g_1^2M_1^2-3g_2^2M_2^2-{16\over 3}g_3^2M_3^2
\Big )\;, \\
{{dM_{u_R}^2}\over {dt}}&=&{2 \over {16\pi ^2}}
\Big (-{16\over 15}g_1^2M_1^2-{16\over 3}g_3^2M_3^2
\Big )\;, \\
{{dM_{d_R}^2}\over {dt}}&=&{2 \over {16\pi ^2}}
\Big (-{4\over 15}g_1^2M_1^2-{16\over 3}g_3^2M_3^2
\Big )\;, \\
{{dM_{l_L}^2}\over {dt}}&=&{2 \over {16\pi ^2}}
\Big (-{3\over 5}g_1^2M_1^2-3g_2^2M_2^2
\Big )\;, \\
{{dM_{e_R}^2}\over {dt}}&=&{2 \over {16\pi ^2}}
\Big (-{12\over 5}g_1^2M_1^2
\Big )\;,
\end{eqnarray}
\end{mathletters}
where
\begin{mathletters}
\begin{eqnarray}
b_i&=&({33\over 5},1,-3) \;, \\
c_i&=&({13\over 15},3,{16\over 3}) \;, \\
c_i^{\prime}&=&({7\over 15},3,{16\over 3}) \;, \\
c_i^{\prime \prime}&=&({9\over 5},3,0) \;, \\
X_t      &=&  M_{Q_L}^2+M_{t _R}^2+M_{H_2}^2+A_t^2\;, \\
X_b      &=&  M_{Q_L}^2+M_{b _R}^2+M_{H_1}^2+A_b^2\;, \\
X_{\tau }&=&  M_{L_L}^2+M_{\tau _R}^2+M_{H_1}^2+A_{\tau }^2\;.
\end{eqnarray}
\end{mathletters}
Here the factors $c_i$, $c_i^{\prime }$, and $c_i^{\prime\prime  }$ are
given by a sum over the fields in the relevant Yukawa coupling,
e.g. $c_i=\sum _f c_i(f)=c_i({H_2})+c_i({Q})+c_i({U^c})$.
The coefficients in front of the gauge coupling parts of Eq.~(41)-(43) can be
understood from the quantum numbers. For the fundamental representations of
$SU(N)$ there is a factor of $(N^2-1)/N$ and for the hypercharge
$U(1)$ one has ${3\over {10}}Y^2$ (with hypercharge suitably normalized,
e.g. $Y_{\tau _R}=2$).

\noindent {\bf One-Loop Effective Potential}

We summarize here the tools needed to construct the one-loop minimization
conditions. The necessary ingredients are the field dependent particle masses;
since we are calculating the tadpole diagrams, we need the particle masses
at the potential
minimum and the Higgs couplings. The tadpoles are calculated in the
$\overline{DR}$ renormalization scheme\cite{dimred}.

 We present here the
contribution from the third generation (s)particles;
the contributions from the other generations
can be obtained with obvious substitutions.
The top and bottom squark and the tau slepton mass matrices
(at the potential minimum) are
\begin{mathletters}
\begin{equation}
\left( \begin{array}{c@{\quad}c}
M_{Q_L}^2 + m_t^2+{1\over 6}
(4M_W^2-M_Z^2)\cos 2\beta & m_t(A_t+\mu \cot \beta )\\
m_t(A_t+\mu \cot \beta ) & M_{t_R}^2 + m_t^2-{2\over 3}(M_W^2-M_Z^2)\cos 2\beta
\end{array} \right)\;,
\end{equation}
\begin{equation}
\left( \begin{array}{c@{\quad}c}
M_{Q_L}^2 + m_b^2-{1\over 6}
(2M_W^2+M_Z^2)\cos 2\beta & m_b(A_b+\mu \tan \beta )\\
m_b(A_b+\mu \tan \beta ) & M_{b_R}^2 + m_b^2+{1\over 3}(M_W^2-M_Z^2)\cos 2\beta
\end{array} \right)\;,\end{equation}
\begin{equation}
\left( \begin{array}{c@{\quad}c}
M_{L_L}^2 + m_{\tau }^2-{1\over 2}(2M_W^2-M_Z^2)\cos 2\beta &
m_{\tau}(A_{\tau}+\mu \tan \beta )\\
m_{\tau }(A_{\tau }+\mu \tan \beta ) & M_{\tau _R}^2
+ m_{\tau }^2+(M_W^2-M_Z^2)\cos 2\beta
\end{array} \right)\;,\end{equation}
\end{mathletters}
which are diagonalized by orthogonal matrices with mixing angles
$\theta _{\tilde{t}}$, $\theta _{\tilde{b}}$, and $\theta _{\tilde{\tau }}$.
The mass eigenstate for the massive third generation sneutrino is
\begin{mathletters}
\begin{eqnarray}
m_{\tilde{\nu}}^2 &=& M_{L_L}^2 + {1\over 2}{M_Z^2}\cos 2\beta\;.
\end{eqnarray}
\end{mathletters}
The relevant Higgs couplings to the squark eigenstates are
\footnotesize
\begin{mathletters}
\begin{eqnarray}
\left \{ \begin{array}{c}
V({\cal J}\tilde{t}_1\tilde{t}_1) \\
V({\cal J}\tilde{t}_2\tilde{t}_2)
\end{array} \right \}
&=&{{igm_t^2}\over {M_W^2}}\pm {{igm_t}\over
{2M_W}}\sin 2\theta _t\left [A_t-\mu \cot \beta \right ]\nonumber \\
&-&{{igM_Z}\over
{\cos \theta _W^{}}}\left [\left \{ \begin{array}{c}
\cos ^2 \theta _t\\
\sin ^2 \theta _t
\end{array} \right \}\left ({1\over 2}
-e_t\sin ^2\theta _W^{}\right )+\left \{ \begin{array}{c}
\sin ^2 \theta _t\\
\cos ^2 \theta _t
\end{array} \right \}(e_t\sin ^2\theta _W^{})
\right ], \nonumber \\ \\
\left \{ \begin{array}{c}
V({\cal J}_{\perp}\tilde{t}_1\tilde{t}_1) \\
V({\cal J}_{\perp}\tilde{t}_2\tilde{t}_2)
\end{array} \right \}
&=&-{{igm_t^2}\over {M_W^2}}\cot \beta
\mp {{igm_t}\over
{2M_W}}\sin 2\theta _t\left [A_t+\mu \tan \beta \right ]\;, \\
\left \{ \begin{array}{c}
V({\cal J}\tilde{b}_1\tilde{b}_1) \\
V({\cal J}\tilde{b}_2\tilde{b}_2)
\end{array} \right \}
&=&{{igm_b^2}\over {M_W^2}}\mp {{igm_b}\over
{2M_W}}\sin 2\theta _b\left [A_b-\mu \tan \beta \right ]\nonumber \\
&+&{{igM_Z}\over
{\cos \theta _W^{}}}\left [\left \{ \begin{array}{c}
\cos ^2 \theta _b\\
\sin ^2 \theta _b
\end{array} \right \}\left ({1\over 2}
+e_b\sin ^2\theta _W^{}\right )+\left \{ \begin{array}{c}
\sin ^2 \theta _b\\
\cos ^2 \theta _b
\end{array} \right \}(e_b\sin ^2\theta _W^{})
\right ], \nonumber \\ \\
\left \{ \begin{array}{c}
V({\cal J}_{\perp}\tilde{b}_1\tilde{b}_1) \\
V({\cal J}_{\perp}\tilde{b}_2\tilde{b}_2)
\end{array} \right \}
&=&-{{igm_b^2}\over {M_W^2}}\tan \beta
\mp {{igm_b}\over
{2M_W}}\sin 2\theta _b\left [A_b\tan \beta+\mu \right ]\;, \\
\left \{ \begin{array}{c}
V({\cal J}\tilde{\tau}_1\tilde{\tau}_1) \\
V({\cal J}\tilde{\tau}_2\tilde{\tau}_2)
\end{array} \right \}
&=&{{igm_\tau^2}\over {M_W^2}}\mp {{igm_\tau}\over
{2M_W}}\sin 2\theta _\tau\left [A_\tau-\mu \tan \beta \right ]\nonumber \\
&+&{{igM_Z}\over
{\cos \theta _W^{}}}\left [\left \{ \begin{array}{c}
\cos ^2 \theta _{\tau}\\
\sin ^2 \theta _{\tau}
\end{array} \right \}\left ({1\over 2}
+e_\tau\sin ^2\theta _W^{}\right )+\left \{ \begin{array}{c}
\sin ^2 \theta _{\tau}\\
\cos ^2 \theta _{\tau}
\end{array} \right \}(e_\tau\sin ^2\theta _W^{})
\right ], \nonumber \\ \\
\left \{ \begin{array}{c}
V({\cal J}_{\perp}\tilde{\tau}_1\tilde{\tau}_1) \\
V({\cal J}_{\perp}\tilde{\tau}_2\tilde{\tau}_2)
\end{array} \right \}
&=&-{{igm_\tau^2}\over {M_W^2}}\tan \beta
\mp {{igm_\tau}\over
{2M_W}}\sin 2\theta _\tau\left [A_\tau\tan \beta+\mu \right ]\;,
\end{eqnarray}
\end{mathletters}
where $e_t$, $e_b$, and $e_\tau$ are the electromagnetic charges $2/3$,
$-1/3$, and $-1$ respectively.
Notice that the D-terms do not contribute to the coupling of
${\cal J}_{\perp}$ to
the squarks.
The mixed couplings (e.g. $V({\cal J}\tilde{t}_1\tilde{t}_2)$)
obviously do not
contribute to the tadpole.
Calculating the tadpole, and making use of the relations
\begin{mathletters}
\begin{eqnarray}
\sin 2\theta _t&=&{{2m_{\tilde{t}_{LR}}^2}\over
{m_{\tilde{t}_1}^2-m_{\tilde{t}_2}^2}}={{2m_t(A_t+\mu \cot \beta )}\over
{m_{\tilde{t}_1}^2-m_{\tilde{t}_2}^2}}\;, \\
\cos 2\theta _t&=&{{m_{\tilde{t}_{L}}^2-m_{\tilde{t}_{R}}^2}\over
{m_{\tilde{t}_1}^2-m_{\tilde{t}_2}^2}}=
{{(M_{Q_L}^2-M_{t_R}^2)+{1\over 6}\cos 2\beta(8M_W^2-5M_Z^2)}\over
{m_{\tilde{t}_1}^2-m_{\tilde{t}_2}^2}}\;, \\
\sin 2\theta _b&=&{{2m_{\tilde{b}_{LR}}^2}\over
{m_{\tilde{b}_1}^2-m_{\tilde{b}_2}^2}}={{2m_b(A_b+\mu \tan \beta )}\over
{m_{\tilde{b}_1}^2-m_{\tilde{b}_2}^2}}\;, \\
\cos 2\theta _b&=&{{m_{\tilde{b}_{L}}^2-m_{\tilde{b}_{R}}^2}\over
{m_{\tilde{b}_1}^2-m_{\tilde{b}_2}^2}}=
{{(M_{Q_L}^2-M_{b_R}^2)-{1\over 6}\cos 2\beta(4M_W^2-M_Z^2)}\over
{m_{\tilde{b}_1}^2-m_{\tilde{\tau }_2}^2}}\;, \\
\sin 2\theta _{\tau}&=&{{2m_{\tilde{\tau}_{LR}}^2}\over
{m_{\tilde{\tau}_1}^2-m_{\tilde{\tau}_2}^2}}={{2m_{\tau}(A_{\tau}+\mu
\tan \beta )}\over
{m_{\tilde{\tau}_1}^2-m_{\tilde{\tau}_2}^2}}\;, \\
\cos 2\theta _\tau&=&{{m_{\tilde{\tau}_{L}}^2-m_{\tilde{\tau}_{R}}^2}\over
{m_{\tilde{\tau}_1}^2-m_{\tilde{\tau}_2}^2}}=
{{(M_{L_L}^2-M_{\tau _R}^2)-{1\over 2}\cos 2\beta(4M_W^2-3M_Z^2)}\over
{m_{\tilde{\tau}_1}^2-m_{\tilde{\tau}_2}^2}}\;,
\end{eqnarray}
\end{mathletters}
one arrives at the top and bottom
quark-squark contribution to the minimization conditions:
\begin{mathletters}
\begin{eqnarray}
\Delta T_1^{(q)}&=&{{3}\over {4\pi ^2v}}\left [m_t^4
\left (\ln {{m_t^2}\over {Q^2}}-1\right )
-m_b^4\left (\ln {{m_b^2}\over {Q^2}}-1\right )\right ]\nonumber \\
&&+{{3}\over {16\pi ^2v}}\Bigg \{m_{\tilde{t}_{1,2}}^2
\left (\ln {{m_{\tilde{t}_{1,2}}^2}\over {Q^2}}-1\right )\Bigg [-2m_t^2
+{1\over 2}M_Z^2\nonumber \\
&&\pm{1\over {m_{\tilde{t}_1}^2-m_{\tilde{t}_2}^2}}\Bigg [{1\over 3}
(8M_W^2-5M_Z^2)\Big [{1\over 2}(M_{Q_L}^2-M_{t_R}^2)+{1\over {12}}\cos 2\beta
(8M_W^2-5M_Z^2)\Big ]\nonumber \\
&&+2m_t^2\Big ((\mu \cot \beta)^2-A_t^2\Big )\Bigg ]
\nonumber \\
&&-m_{\tilde{b}_{1,2}}^2
\left (\ln {{m_{\tilde{b}_{1,2}}^2}\over {Q^2}}-1\right )\Bigg [-2m_b^2
+{1\over 2}M_Z^2\nonumber \\
&&\pm{1\over {m_{\tilde{b}_1}^2-m_{\tilde{b}_2}^2}}\Bigg [{1\over 3}
(4M_W^2-M_Z^2)\Big [{1\over 2}(M_{Q_L}^2-M_{b_R}^2)-{1\over {12}}\cos 2\beta
(4M_W^2-M_Z^2)\Big ]\nonumber \\
&&+2m_b^2\Big ((\mu \tan \beta)^2-A_b^2\Big )\Bigg ]
\Bigg \}\;,
\\
\Delta T_2^{(q)}&=&-{{3}
\over {4\pi ^2v}}\left [m_t^4
\left (\ln {{m_t^2}\over {Q^2}}-1\right )\cot \beta +m_b^4
\left (\ln {{m_b^2}\over {Q^2}}-1\right )\tan \beta \right ]\nonumber \\
&+&{{3}\over {8\pi ^2v}}\Bigg \{m_{\tilde{t}_{1,2}}^2
\left (\ln {{m_{\tilde{t}_{1,2}}^2}\over {Q^2}}-1\right )\cot \beta
\Bigg [m_t^2
\pm{1\over {m_{\tilde{t}_1}^2-m_{\tilde{t}_2}^2}}
m_t^2(A_t+\mu \cot \beta )(A_t+\mu \tan \beta )\Bigg ]\nonumber \\
&+&m_{\tilde{b}_{1,2}}^2
\left (\ln {{m_{\tilde{b}_{1,2}}^2}\over {Q^2}}-1\right )\tan \beta
\Bigg [m_b^2
\pm{1\over {m_{\tilde{b}_1}^2-m_{\tilde{b}_2}^2}}
m_b^2(A_b+\mu \cot \beta )(A_b+\mu \tan \beta )\Bigg ]\Bigg \}.
\end{eqnarray}
\end{mathletters}
Also the tau lepton-slepton and sneutrino contributions are
\begin{mathletters}
\begin{eqnarray}
\Delta T_1^{(l)}&=&{{1}\over {4\pi ^2v}}\left [-m_\tau^4\left (\ln
{{m_\tau^2}\over {Q^2}}-1\right )\right ]\nonumber \\
&&+{{1}\over {16\pi ^2v}}\Bigg \{m_{\tilde{\nu}}^2{M_Z}^2
\left (\ln {{m_{\tilde{\nu}}^2}\over {Q^2}}-1\right )\nonumber \\
&&-m_{\tilde{\tau}_{1,2}}^2
\left (\ln {{m_{\tilde{\tau}_{1,2}}^2}\over {Q^2}}-1\right )\Bigg [-2m_\tau^2
+{1\over 2}M_Z^2\nonumber \\
&&\pm{1\over {m_{\tilde{\tau}_1}^2-m_{\tilde{\tau}_2}^2}}\Bigg [
(4M_W^2-3M_Z^2)\Big [{1\over 2}(M_{L_L}^2-M_{\tau _R}^2)
-{1\over {4}}\cos 2\beta
(4M_W^2-3M_Z^2)\Big ]\nonumber \\
&&+2m_\tau^2\Big ((\mu \tan \beta)^2-A_\tau^2\Big )\Bigg ]
\Bigg \}\;,
\\
\Delta T_2^{(l)}&=&-{{1}
\over {4\pi ^2v}}\left [m_\tau^4
\left (\ln {{m_\tau^2}\over {Q^2}}-1\right )\tan \beta \right ]
+{{1}\over {8\pi ^2v}}\Bigg \{m_{\tilde{\tau}_{1,2}}^2
\left (\ln {{m_{\tilde{\tau}_{1,2}}^2}\over {Q^2}}-1\right )\tan \beta
\nonumber \\
&&\times \Bigg [m_\tau^2
\pm{1\over {m_{\tilde{\tau}_1}^2-m_{\tilde{\tau}_2}^2}}
m_\tau^2(A_\tau+\mu \cot \beta )(A_\tau+\mu \tan \beta )\Bigg ]\Bigg \}\;,
\end{eqnarray}
\end{mathletters}
There are similar contributions from the first and second generations.
Most of these terms in $\Delta T_1$ and all of the terms in $\Delta T_2$
are proportional to some powers of the
quark or lepton mass, which is negligible in the light generations.
However, there exist contributions to $\Delta T_1$ which
are proportional to $M_Z^2$ and $M_W^2$;
these are zero only in the limit in which
the squarks (or sleptons) are degenerate within each generation.
This is not necessarily a good approximation; we find that
the light squark/slepton contribution can be larger than
the gauge boson contribution (see below), especially for moderate or large
values of $m_0^{}$ and $m_{1\over 2}$. It is therefore important
to include the light squarks and sleptons in a full one-loop analysis.
Explicitly, the light squark and slepton contribution is
\begin{mathletters}
\begin{eqnarray}
\Delta T_1^{(lq)}&=&{{(2)3}\over {16\pi ^2v}}\Bigg \{m_{\tilde{u}_{1,2}}^2
\left (\ln {{m_{\tilde{u}_{1,2}}^2}\over {Q^2}}-1\right )
\Bigg [{1\over 2}M_Z^2\pm ({1\over 2}){1\over 3}(8M_W^2-5M_Z^2)\Bigg ]
\nonumber \\
&&-m_{\tilde{d}_{1,2}}^2
\left (\ln {{m_{\tilde{d}_{1,2}}^2}\over {Q^2}}-1\right )
\Bigg [{1\over 2}M_Z^2\pm({1\over 2}){1\over 3}(4M_W^2-M_Z^2)\Bigg ]
\Bigg \}\;,
\\
\Delta T_1^{(ll)}&=&{{(2)1}\over {16\pi ^2v}}\Bigg \{m_{\tilde{\nu}}^2
\left (\ln {{m_{\tilde{\nu}}^2}\over {Q^2}}-1\right )
\Bigg [M_Z^2 \Bigg ]
\nonumber \\
&&-m_{\tilde{e}_{1,2}}^2
\left (\ln {{m_{\tilde{e}_{1,2}}^2}\over {Q^2}}-1\right )
\Bigg [{1\over 2}M_Z^2\pm({1\over 2})(4M_W^2-3M_Z^2)\Bigg ]
\Bigg \}\;,
\\
\Delta T_2^{(lq)}&=&0\;,
\Delta T_2^{(ll)}=0\;,
\end{eqnarray}
\end{mathletters}
where the factor of two includes both light generations.

If we neglect the contribution from the bottom quark and from the D-term
contributions to the squark masses and couplings, the equations above
reduce to
\begin{mathletters}
\begin{eqnarray}
\Delta T_1^{(q)}&=&{{3m_t^2}\over {8\pi ^2v}}\Bigg [2f(m_t^2)
-f(m_{\tilde{t}_{1}}^2)-f(m_{\tilde{t}_{2}}^2)+
{{f(m_{\tilde{t}_{1}}^2)-f(m_{\tilde{t}_{2}}^2)}
\over {m_{\tilde{t}_1}^2-m_{\tilde{t}_2}^2}}
\Big ((\mu \cot \beta)^2-A_t^2\Big )\Bigg ]
\;, \\
\Delta T_2^{(q)}&=&-{{3m_t^2\cot \beta }\over {8\pi ^2v}}\Bigg [2f(m_t^2)
-f(m_{\tilde{t}_{1}}^2)-f(m_{\tilde{t}_{2}}^2)\nonumber \\
&&-{{f(m_{\tilde{t}_{1}}^2)-f(m_{\tilde{t}_{2}}^2)}
\over {m_{\tilde{t}_1}^2-m_{\tilde{t}_2}^2}}
(A_t+\mu \cot \beta )(A_t+\mu \tan \beta )\Bigg ]\;,
\end{eqnarray}
\end{mathletters}
where
\begin{eqnarray}
f(m^2)=m^2\left (\ln {{m^2}\over {Q^2}}-1\right )\;.
\end{eqnarray}

The neutralino mass matrix is
\begin{equation}
M_N=\left( \begin{array}{c@{\quad}c@{\quad}c@{\quad}c}
M_1 & 0 & -M_Z\cos \beta \sin \theta _W^{} & M_Z\sin \beta \sin \theta _W^{}
\\
0 & M_2 & M_Z\cos \beta \cos \theta _W^{} & -M_Z\sin \beta \cos \theta _W^{}
\\
-M_Z\cos \beta \sin \theta _W^{} & M_Z\cos \beta \cos \theta _W^{} & 0 & \mu
\\
M_Z\sin \beta \sin \theta _W^{} & -M_Z\sin \beta \cos \theta _W^{} & \mu & 0
\end{array} \right)\;.
\end{equation}
This mass matrix is symmetric and can be diagonalized by a single matrix $Z$
as\cite{hk}
\begin{eqnarray}
&&Z^*M_NZ^{-1}\;,
\end{eqnarray}
We choose $Z$ to be a real matrix; then the diagonalized neutrino mass matrix
can have negative entries. We let the entries be $\epsilon _iM_{\chi _i^0}$
where $M_{\chi _i^0}$ are {\it positive} masses and $\epsilon _i$ takes on a
value of $+1$ or $-1$.
The diagonalization can be done numerically, or one can use the analytic
expressions\cite{eksa}
\begin{mathletters}
\begin{eqnarray}
\epsilon _1M_{\chi _1^0}&=&-({1\over 2}a-{1\over 6}C_2)^{1/2}
+\left [-{1\over 2}a-{1\over 3}C_2+{{C_3}\over {(8a-{8\over 3}C_2)^{1/2}}}
\right ]^{1/2}+{1\over 4}(M_1+M_2)\;, \nonumber \\ \\
\epsilon _2M_{\chi _2^0}&=&+({1\over 2}a-{1\over 6}C_2)^{1/2}
-\left [-{1\over 2}a-{1\over 3}C_2-{{C_3}\over {(8a-{8\over 3}C_2)^{1/2}}}
\right ]^{1/2}+{1\over 4}(M_1+M_2)\;, \nonumber \\ \\
\epsilon _3M_{\chi _3^0}&=&-({1\over 2}a-{1\over 6}C_2)^{1/2}
-\left [-{1\over 2}a-{1\over 3}C_2+{{C_3}\over {(8a-{8\over 3}C_2)^{1/2}}}
\right ]^{1/2}+{1\over 4}(M_1+M_2)\;, \nonumber \\ \\
\epsilon _4M_{\chi _4^0}&=&+({1\over 2}a-{1\over 6}C_2)^{1/2}
+\left [-{1\over 2}a-{1\over 3}C_2-{{C_3}\over {(8a-{8\over 3}C_2)^{1/2}}}
\right ]^{1/2}+{1\over 4}(M_1+M_2)\;, \nonumber \\
\end{eqnarray}
\end{mathletters}
where
\begin{mathletters}
\begin{eqnarray}
C_2&=&(M_1M_2-M_Z^2-\mu ^2)-{3\over 8}(M_1+M_2)^2\;, \\
C_3&=&-{1\over 8}(M_1+M_2)^3
+{1\over 2}(M_1+M_2)(M_1M_2-M_Z^2-\mu^2)+(M_1+M_2)\mu ^2
\nonumber \\
&&+(M_1\cos ^2\theta _W^{}
+M_2\sin ^2\theta _W^{})M_Z^2+\mu M_Z^2\sin 2\beta\;, \\
C_4&=&-(M_1\cos ^2\theta _W^{}+M_2\sin ^2\theta _W^{})
M_Z^2\mu\sin 2\beta -M_1M_2\mu^2\nonumber \\
&&+{1\over 4}(M_1+M_2)
[(M_1+M_2)\mu ^2+(M_1\cos ^2\theta _W^{}+M_2\sin ^2\theta _W^{})
M_Z^2+\mu M_Z^2\sin 2\beta] \nonumber \\
&&+{1\over 16}(M_1M_2-M_Z^2-\mu^2)(M_1+M_2)^2
-{3\over 256}(M_1+M_2)^4\;, \\
a&=&{1\over {2^{1/3}}}{\rm Re}\left [-S+i(D/27)^{1/2}\right ]^{1/3}\;, \\
D&=&-4U^3-27S^2\;, \quad U=-{1\over 3}C_2^2-4C_4, \quad S=-C_3^2
-{2\over 27}C_2^3  +{8\over 3}C_2C_4\;.
\end{eqnarray}
\end{mathletters}
These masses given by the above expression are not necessarily such that
$M_{\chi_1^0}<M_{\chi_2^0}<M_{\chi_3^0}<M_{\chi_4^0}$, but the eigenstates
can be relabeled.
We have corrected a typographical error in the definition of $U$ given in
Ref.~\cite{eksa}.
The contribution to the minimization conditions is
\begin{mathletters}
\begin{eqnarray}
\Delta T_1^{(\chi ^0)}&=&-{1\over 2}\sum_{i=1}^4{{gM_{\chi _i^0}^3}
\over {4\pi ^2}}\Bigg [Q_{ii}^{\prime \prime }\cos \beta +
S_{ii}^{\prime \prime }
\sin \beta \Bigg ]\Bigg (\ln {{M_{\chi _i^0}^2}
\over {Q^2}}-1\Bigg )\;, \\
\Delta T_2^{(\chi ^0)}&=&-{1\over 2}\sum_{i=1}^4{{gM_{\chi _i^0}^3}
\over {4\pi ^2}}\Bigg [Q_{ii}^{\prime \prime }
\sin \beta -
S_{ii}^{\prime \prime }
\cos \beta \Bigg ]\Bigg (\ln {{M_{\chi _i^0}^2}
\over {Q^2}}-1\Bigg )\;.
\end{eqnarray}
\end{mathletters}
The factors $Q_{ii}^{\prime \prime }$ and $S_{ii}^{\prime \prime }$ are
defined as\cite{hunters}
\begin{mathletters}
\begin{eqnarray}
Q_{ii}^{\prime \prime }&=&\left [Z_{i3}(Z_{i2}-Z_{i1}\tan \theta _w)
\right ]\epsilon _i \;, \\
S_{ii}^{\prime \prime }&=&\left [Z_{i4}(Z_{i2}-Z_{i1}\tan \theta _w)
\right ]\epsilon _i \;,
\end{eqnarray}
\end{mathletters}
where $\epsilon _i$ is the sign of the $i$th eigenvalue of the neutralino
mass matrix.
The mixing matrix $Z$ can also be given by analytic expressions\cite{eksa}
\begin{mathletters}
\begin{eqnarray}
{{Z_{i2}}\over {Z_{i1}}}&=&-{1\over {\tan \theta _W^{}}}
{{M_1-\epsilon _iM_{\chi _i^0}}\over {M_2-\epsilon _iM_{\chi _i^0}}}\;,
\\
{{Z_{i3}}\over {Z_{i1}}}&=&{{-\mu [M_2-\epsilon _iM_{\chi _i^0}]
[M_1-\epsilon _iM_{\chi _i^0}]-M_Z^2\sin \beta \cos \beta
[(M_1-M_2)\cos ^2\theta _W^{}+M_2-\epsilon _iM_{\chi _i^0}]}\over
{M_Z[M_2-\epsilon _iM_{\chi _i^0}]\sin \theta _W^{}[-\mu \cos \beta+
\epsilon _iM_{\chi _i^0} \sin \beta ]}}\;, \nonumber \\ \\
{{Z_{i4}}\over {Z_{i1}}}&=&{{-\epsilon _iM_{\chi _i^0}
[M_2-\epsilon _iM_{\chi _i^0}]
[M_1-\epsilon _iM_{\chi _i^0}]-M_Z^2\cos ^2\beta
[(M_1-M_2)\cos ^2\theta _W^{}+M_2-\epsilon _iM_{\chi _i^0}]}\over
{M_Z[M_2-\epsilon _iM_{\chi _i^0}]\sin \theta _W^{}[-\mu \cos \beta+
\epsilon _iM_{\chi _i^0} \sin \beta ]}}\;, \nonumber \\
\end{eqnarray}
\end{mathletters}
and
\begin{eqnarray}
Z_{i1}=\left [1+\left ({{Z_{i2}}\over {Z_{i1}}}\right )^2
+\left ({{Z_{i3}}\over {Z_{i1}}}\right )^2+\left ({{Z_{i4}}
\over {Z_{i1}}}\right )^2
\right ]^{-1/2}\;.
\end{eqnarray}
In terms of the mixing matrix $Z$ the bino and gaugino purities are defined
as
\begin{mathletters}
\begin{eqnarray}
B_P&=&Z_{11}^2\;, \\
G_P&=&Z_{11}^2+Z_{12}^2\;,
\end{eqnarray}
\end{mathletters}
respectively.

The chargino mass matrix is
\begin{equation}
M_C=\left( \begin{array}{c@{\quad}c}
M_2 & \sqrt{2}M_W\sin \beta \\
\sqrt{2}M_W\cos \beta & -\mu
\end{array} \right)\;.
\end{equation}
This mass matrix is not symmetric and must be diagonalized by two matrices
$U$ and $V$ as\cite{hk}
\begin{eqnarray}
&&U^*M_CV^{-1}\;,
\end{eqnarray}
where
\begin{eqnarray}
U&=&O_-\;,\;\;\;\;V=\left \{ \begin{array}{lcr}
O_+&,&\det X\geq 0 \\
\sigma _3O_+&,&\det X<0
\end{array}\right .\;,\;\;\;\;
O_{\pm}=\left( \begin{array}{c@{\quad}c}
\cos \phi _{\pm} & \sin \phi _{\pm} \\
-\sin \phi _{\pm} & \cos \phi _{\pm}\end{array} \right)\;.
\end{eqnarray}
Here $\sigma _3$ is the Pauli matrix, and
\begin{eqnarray}
\tan 2\phi _-&=&2\sqrt{2}M_W{{-\mu \sin \beta + M_2\cos \beta }\over
{M_2^2-\mu ^2-2M_W^2\cos 2\beta }}\;, \\
\tan 2\phi _+&=&2\sqrt{2}M_W{{-\mu \cos \beta + M_2\sin \beta }\over
{M_2^2-\mu ^2+2M_W^2\cos 2\beta }}\;.
\end{eqnarray}
The chargino masses are
\begin{eqnarray}
M_{\chi ^{\pm }}^2&=&
{1\over 2}\Bigg [M_2^2+\mu ^2+2M_W^2 \nonumber \\
&&\pm \left [(M_2^2-\mu ^2)^2
+4M_W^4\cos ^22\beta+4M_W^2(M_2^2+\mu ^2-2M_2\mu \sin 2\beta )\right ]^{1/2}
\Bigg ]\;.
\end{eqnarray}
The contribution to the minimization conditions is
\begin{mathletters}
\begin{eqnarray}
\Delta T_1^{(\chi ^{\pm })}&=&-\sum_{i=1}^2{{gM_{\chi _i^{\pm }}^3}
\over {4\pi ^2}}\Bigg [Q_{ii}\cos \beta -
S_{ii}\sin \beta \Bigg ]
\Bigg (\ln {{M_{\chi _i^{\pm }}^2}
\over {Q^2}}-1\Bigg )\;, \\
\Delta T_2^{(\chi ^{\pm })}&=&-\sum_{i=1}^2{{gM_{\chi _i^{\pm }}^3}
\over {4\pi ^2}}\Bigg [Q_{ii}\sin \beta +
S_{ii}\cos \beta \Bigg ]
\Bigg (\ln {{M_{\chi _i^{\pm }}^2}
\over {Q^2}}-1\Bigg )\;.
\end{eqnarray}
\end{mathletters}
The factors $Q_{ii}$ and $S_{ii}$ are defined as
\begin{mathletters}
\begin{eqnarray}
Q_{ii}&=&\sqrt{1\over 2}V_{i1}U_{i2}\;, \\
S_{ii}&=&\sqrt{1\over 2}V_{i2}U_{i1}\;.
\end{eqnarray}
\end{mathletters}

The Higgs bosons and Goldstone bosons contribute the following contributions
in the Landau gauge
\begin{mathletters}
\begin{eqnarray}
\Delta T_1^{(H)}&=&{{gM_{H^{\pm }}^2}\over {32\pi^2}}
\left (2M_W-{{M_Z}\over {\cos \theta _w}}\right )\cos 2\beta
\left (\ln {{M_{H^{\pm }}^2}\over {Q^2}}-1\right )\nonumber \\
&&+{{gM_ZM_h^2}\over {64\pi^2\cos \theta _w}}(-2\cos 2\alpha
+\cos 2\beta )
\left (\ln {{M_h^2}\over {Q^2}}-1\right )\nonumber \\
&&+{{gM_ZM_H^2}\over {64\pi^2\cos \theta _w}}(2\cos 2\alpha
+\cos 2\beta )
\left (\ln {{M_H^2}\over {Q^2}}-1\right )\nonumber \\
&&-{{gM_ZM_A^2}\over {64\pi^2\cos \theta _w}}\cos 2\beta
\left (\ln {{M_A^2}\over {Q^2}}-1\right )\;, \\
\Delta T_2^{(H)}&=&{{gM_WM_{H^{\pm }}^2}\over {16\pi^2}}\sin 2\beta
\left (\ln {{M_{H^{\pm }}^2}\over {Q^2}}-1\right )\nonumber \\
&&+{{gM_ZM_h^2}\over {64\pi^2\cos \theta _w}}(\sin 2\alpha +\sin 2\beta)
\left (\ln {{M_h^2}\over {Q^2}}-1\right )\nonumber \\
&&+{{gM_ZM_H^2}\over {64\pi^2\cos \theta _w}}(-\sin 2\alpha +\sin 2\beta)
\left (\ln {{M_H^2}\over {Q^2}}-1\right )\;.
\end{eqnarray}
\end{mathletters}
The angle factor $\alpha $ can be eliminated in the above equations using
the tree level relations for the Higgs masses:
\begin{mathletters}
\begin{eqnarray}
-2\cos 2\alpha + \cos 2\beta &=&\cos 2\beta\left ({{3M_H^2+M_h^2-4M_Z^2}\over
{M_H^2-M_h^2}}\right )\;, \\
2\cos 2\alpha + \cos 2\beta &=&\cos 2\beta\left ({{3M_h^2+M_H^2-4M_Z^2}\over
{M_h^2-M_H^2}}\right )\;, \\
\sin 2\alpha +\sin 2\beta &=&\sin 2\beta\left ({{2M_h^2}\over {M_h^2-M_H^2}}
\right )\;, \\
-\sin 2\alpha +\sin 2\beta &=&\sin 2\beta\left ({{2M_H^2}\over {M_H^2-M_h^2}}
\right )\;.
\end{eqnarray}
\end{mathletters}
The gauge boson contribution is
\begin{mathletters}
\begin{eqnarray}
\Delta T_1^{(GB)}&=&{{3gM_W^3}\over {16\pi^2}}\cos 2\beta
\left (\ln {{M_W^2}\over {Q^2}}-1\right )\nonumber \\
&&+{{3gM_Z^3}\over {32\pi^2\cos \theta _w}}\cos 2\beta
\left (\ln {{M_Z^2}\over {Q^2}}-1\right )\;, \\
\Delta T_2^{(GB)}&=&{{3gM_W^3}\over {16\pi^2}}\sin 2\beta
\left (\ln {{M_W^2}\over {Q^2}}-1\right )\nonumber \\
&&+{{3gM_Z^3}\over {32\pi^2\cos \theta _w}}\sin 2\beta
\left (\ln {{M_Z^2}\over {Q^2}}-1\right )\;.
\end{eqnarray}
\end{mathletters}

Then the minimization conditions at one-loop are the following:
\begin{mathletters}
\begin{eqnarray}
T_1+\sum _i\Delta T_1^{(i)}&=&0\;, \label{t1p} \\
T_2+\sum _i\Delta T_2^{(i)}&=&0\;, \label{t2p}
\end{eqnarray}
\end{mathletters}
where $i=q,l,lq,ll,\chi ^0,\chi ^{\pm},H,GB$.

\end{document}